\documentclass[12pt]{article}
\usepackage[latin9]{inputenc}
\usepackage{geometry}
\usepackage{float}
\usepackage{booktabs}
\usepackage{amsmath}
\usepackage{amssymb}
\usepackage{stmaryrd}
\usepackage{graphicx}
\usepackage{setspace}
\usepackage[authoryear]{natbib}
\usepackage[unicode=true,
 bookmarks=false,
 breaklinks=false,pdfborder={0 0 1},backref=section,colorlinks=false]
 {hyperref}
\hypersetup{
 colorlinks,citecolor=blue,pdftex}

\makeatletter

\providecommand{\tabularnewline}{\\}

\usepackage{amsfonts}
\usepackage{amsthm}
\usepackage{lscape}
\usepackage{appendix}
\usepackage{dcolumn}
\usepackage{rotating}
\usepackage{comment}
\usepackage{color}

\usepackage{pgf}
\usepackage{tikz}\usepackage{caption}
\usepackage{subcaption}
\usepackage{tkz-tab}
\usepackage{tikz-3dplot}
\usetikzlibrary{calc}
\usetikzlibrary{arrows, automata}

\newcolumntype{d}[1]{D{.}{.}{#1}}
\newcolumntype{t}[1]{D{,}{,}{#1}}
\newcolumntype{i}[1]{D{.}{}{#1}}

\theoremstyle{plain} 

\usepackage{authblk}

\author[1]{Sukjin Han}
\author[2]{Eric Schulman}
\author[3]{$$\vspace{-1.8cm}$$Kristen Grauman}
\author[3]{Santhosh Ramakrishnan}
\affil[1]{School of Economics, University of Bristol}
\affil[2]{U.S. Congressional Budget Office}
\affil[3]{Department of Computer Science, University of Texas at Austin}
\date{}                     
\makeatother

\begin{document}
\title{Shapes as Product Differentiation\thanks{For helpful discussions, the authors are grateful to Jorge Balat,
Aureo de Paula, Phil Haile, Greg Howard, Alessandro Iaria, Roger Moon,
Jesse Shapiro, Matt Shum, Haeyeon Yoon, participants in the 2022 North
American Summer Meeting, the 2021 North American Winter Meeting, the
2020 European Winter Meeting, the 2020 North American Winter Meeting
of the Econometric Society, the 2022 IAAE Conference, the 25th Texas
Econometrics Camp, and the KAEA Virtual Seminar Series. We acknowledge
the generous provision of the data set by Monotype Inc. Sukjin especially
thanks Nathan Ford, the Customer Analytics \& Insight Director of
Monotype, for fruitful conversations throughout the project, and graphic
designers, Jaewon Seok and Jiwon Park, for early discussions on font
design and font markets. Lastly, we appreciate research assistants,
George Kyungho Lee and Po-Yuan Huang, for their excellent work. The
views expressed in this paper are the authors' and should not be interpreted
as those of CBO. Corresponding author: Sukjin Han \protect\href{mailto:sukjin.han@gmail.com}{sukjin.han@gmail.com}}}
\date{This Draft: \today{\small{}\vspace{-0.5cm}
}}
\maketitle
\begin{abstract}
Many differentiated products have key attributes that are \textit{unstructured}
and thus high-dimensional (e.g., design, text). Instead of treating
unstructured attributes as unobservables in economic models, quantifying
them can be important to answer interesting economic questions. To
propose an analytical framework for these types of products, this
paper considers one of the simplest design products---fonts---and
investigates merger and product differentiation using an original
dataset from the world's largest online marketplace for fonts. We
quantify font shapes by constructing embeddings from a deep convolutional
neural network. Each embedding maps a font's shape onto a low-dimensional
vector. In the resulting product space, designers are assumed to engage
in Hotelling-type spatial competition. From the image embeddings,
we construct two alternative measures that capture the degree of design
differentiation. We then study the causal effects of a merger on the
merging firm's creative decisions using the constructed measures in
a synthetic control method. We find that the merger causes the merging
firm to increase the visual variety of font design. Notably, such
effects are not captured when using traditional measures for product
offerings (e.g., specifications and the number of products) constructed
from structured data.

\vspace{0.1in}



\noindent 
\end{abstract}

\section{Introduction\label{sec:Introduction}}

Many differentiated products considered in economic analyses have
important attributes that are \textit{unstructured}. Examples include
design elements in products such as automobiles, houses, furniture,
and clothing or digital products such as mobile applications. Other
obvious examples include creative features in books, music, movies,
and fine arts. Unstructured attributes in these products are typically
in visual or textual forms and thus are high-dimensional. More generally,
products well beyond these categories are often presented to consumers
in visual and textual forms: for example, product packages in supermarkets
and online catalogs in e-commerce (e.g., Amazon, Airbnb, Yelp, Zillow).
These attributes are one of the first pieces of information consumers
receive along with more structured attributes such as price and product
specifications. As a result, unstructured attributes are important
decision factors for consumers and thus are key decision variables
for producers. 

Economists are aware of the role of unstructured attributes. Product
attributes are an important component of economic models, such as
discrete-choice models (\citet{mcfadden1973conditional}, \citet{berry1995automobile})
and hedonic models (\citet{rosen1974hedonic}, \citet{bajari2005demand}).
These models treat product attributes as both low-dimensional observable
variables and a (typically scalar) unobservable variable. In these
models, the scalar unobservable variable normally captures the high-dimensional,
unstructured attributes, including design and other original features
of the products.

Although this tradition has its own merits, certain economic questions
are better answered by treating unstructured attributes as observables.
For example, one may ask how the style of products evolves over time---that
is, how the fashion changes---in accordance with market conditions.
One may further ask how product differentiation in this creative dimension
affects the product\textquoteright s market power or is affected by
market shocks such as vertical integration. These questions entail
the issues of measurability and dimensionality of unstructured attributes.
Due to these challenges, there has been little understanding about
the production of creative attributes---namely, product differentiation
decisions---in the realm of quantitative economics.\footnote{\citet{galenson2000age,galenson2001creating} study artists and their
career choices and paths in fine arts. Their main approach to quantitative
analyses is to use price as a proxy to measure the value of artworks.} 

In this paper, we propose a framework to quantify the design-oriented
attributes of a product and construct a low-dimensional space of products
and measures for the degree of product differentiation using the Euclidean
distance endowed in the space. We then illustrate how some of the
economic questions above can be answered using this framework, such
as market-driven product differentiation decisions of product designers
in multi-product firms.

For the purpose of this paper, we consider a particular design product:
fonts. We use a dataset obtained from the world's largest online marketplace
for Roman alphabet fonts. There are a few reasons to study the market
for fonts. First, font is one of the simplest visually differentiated
products. The shapes (i.e., two-dimensional monochrome visual information)
of a fixed number of characters mostly describe the product.\footnote{These shapes are called typefaces.}
This visual simplicity greatly facilitates our analysis. Second, the
visual information is simple to understand but important in predicting
the functionality and value of the product. Third, fonts are ubiquitous
products; thus, the market for fonts is large with frequent productions
and transactions. The online marketplace we consider is the world's
largest and has over 28,000 fonts (produced by font design firms called
foundries) and 2,400,000 transactions over the past six years. Fourth,
interesting policies are involved in this market, such as vertical
integration. Finally and most importantly, font is a stylized product
that saliently captures a key aspect many products in the market have
in common: design attributes.

The main challenge in quantitatively analyzing the market for fonts
is that the main product attributes, their shapes, are high-dimensional.
To address this challenge, we represent font shapes as low-dimensional
neural network embeddings and construct a corresponding space of fonts.
In particular, we adapt a state-of-the-art method in convolutional
neural networks (\citet{wang2014triplet}, \citet{schroff2015facenet}),
where the network directly learns to map font images to a compact
Euclidean space---that is, the embedding space---in which perceived
visual similarity is preserved. The algorithm is sophisticated enough
to recognize the \textit{style} of font shapes, which is crucial for
our purpose. 

Another challenge is to ensure that the resulting embeddings represent
economic agents' perceptions of the product's visual information.
To this end, we employ two strategies. First, we use the images of
entire \textit{pangrams} (instead of individual alphabet letters)
as inputs in the neural network.\footnote{A pangram is a sentence that contains all the alphabet letters.}
Because pangrams effectively capture important design elements that
cannot be seen in individual letters (e.g., spacing, deep-height,
up-height, and ligature), they are the most relevant decision variables
for font designers and consumers. Second, we demonstrate that the
obtained image embeddings contains a substantial amount of information
that is mutually shared with \textit{tags}, which are word phrases
that describe fonts (e.g., ``curly,'' ``flowing,'' ``geometric,''
``organic''). Tags are assigned to each font by font designers and
consumers and thus, we believe, represent the economic agents' perceptions
of the product. We construct word embeddings from these tags using
a simple neural network and calculate the mutual information between
the word and image embeddings.

Why do we consider a convolutional neural network? Font shapes involve
a non-linear interaction between many neighboring pixels. Considering
individual pixels separately or recognizing interactions in a restrictive
model provides little information about the overall shape of a font.
By considering how neighboring pixels interact in a very flexible
model, the deep neural network outperforms other machine learning
methods such as LASSO, random forest, and boosting that use pixels
or other hand-designed features (edges, corners, etc.); see \citet{Goodfellow-et-al-2016}.
In particular, the deep convolutional neural network is designed to
effectively capture the spatial correlation between nearby pixels.
Although neural networks are generally known to be less interpretable
(\citet{friedman2001elements}) than other learning methods due to
model flexibility,\footnote{One approach to overcome this is to consider ``hand-crafted features.''
However, feature selection can generally be arbitrary and there can
be arbitrarily many possible features, which hinders the interpretation.} we show how an interpretable embedding space can be learned through
visual similarity. Most importantly, instead of attempting to interpret
each embedding value, our approach is to give meanings to the distance
metric of the embedding space and subsequently construct the product
differentiation measures based on it.

Given the embedding space, a font designer's decision of a typeface
design is equivalent to choosing a location in the space. This location
choice is a strategic decision that depends on the choices of other
designers in the space. In this sense, the abstract space of fonts
we construct can be viewed as a location-analog model (\citet{hotelling1929})
with Lancasterian characteristics (\citet{lancaster1966new,lancaster1971consumer}).
Based on this space, we construct two alternative differentiation
measures using the image embeddings, namely a distance to Averia (i.e.,
the average font) and a gravity measure, which succinctly represent
the location choice of a designer relative to others' choices.

To illustrate the usefulness of our approach, we conduct a causal
analysis of how a merger affected the merged firm's design decisions
before and after the merger. In June 2014, a major font foundry was
acquired by the company that owns the online marketplace. Before the
merger, the merging firm was selling fonts as a third party and competing
with the foundries owned by the marketplace. The main motivation of
the analysis is that designers are not only artists but also economic
agents who are affected by market conditions. We use the constructed
design differentiation measures as the main dependent variables. To
estimate the effect of merger on differentiation, we use the synthetic
control method (\citet{abadie2003economic}, \citet{abadie2010synthetic}).
We employ this method because, although there is only one treated
(i.e., merged) foundry, we can construct a suitable weighted average
of a comparison group from untreated foundries. We provide arguments
why strategic spillovers may be weak in our setting by showing that
each control foundry produces substantially different products than
the treated foundry, even though the synthetic control behaves similarly
as the treated before the merger. The latter is achieved due to the
rich information contained in the embeddings, which serve as the main
predictors in constructing the synthetic control.

Our main finding is that, relative to the synthetic control, the merged
foundry produced fonts with greater visual variety after the merger
and that this effect was statistically significant. One of the explanations
is that the degree of product differentiation may have increased after
the merger to avoid cannibalization. Notably, we find that such effects
are not captured when traditional measures for product offerings are
used from structured data (i.e., the number of products and specifications;
\citet{berry2001}). This illustrates the importance of more sophisticated
product offerings measures as employed in the current paper.

\subsection{Contributions and Related Literature}

\subsubsection*{Machine Learning and Social Science Research}

To our knowledge, this is one of the first few papers using neural
network embeddings for visual data in the economic analysis of markets
and industries. Earlier work that analyzes visual data uses methods
that are partly or fully human-aided. \citet{glaeser2018big} use
visual data from Google Street View to predict the economic prosperity
of neighborhoods, but they assign scores to street images based on
human surveys on the visual perception of street quality and safety.
\citet{gross2016creativity} investigates how competition influences
creative production in a commercial logo design competition. Whereas
\citet{gross2016creativity} uses hand-crafted features to create
a perceptual hash code for comparing images, we learn image embeddings
based on recent advances in deep learning that have been shown to
work extremely well for high-dimensional image data (\citet{krizhevsky2012imagenet},
\citet{simonyan2014very}, \citet{he2016deep}). Deep learning methods
are used in more recent studies. \citet{zhang2017much} show how the
quality and specific attributes of property images on Airbnb can affect
the demand. The quality and attributes are human-labeled, and a convolutional
neural network is used to train the images based on the labels. Our
approach does not require subjective human labeling. As independent
and contemporaneous work, \citet{bajari2018hedonic} estimate a hedonic
function for apparel consumption using a deep neural network based
on visual and textual data. We also utilize textual data but in such
a way that gives economic meanings to the image embeddings we obtain.
The most significant difference between our approach and the two aforementioned
studies is that we consider images as the main response variables
of market shocks and construct differentiation measures based on neural
network embeddings. Using these measures also helps gain interpretability
and robustness in the quantities resulting from the network training.
In addition, we use a different neural network algorithm based on
\textit{image triplets} that is suitable to our setting. Another recent
work by \citet{magnolfi2022triplet} considers triplet embeddings
to characterize the product space for demand estimation. Although
we share a similar motivation, their embeddings are computed after
human making comparisons of product triplets while ours are elicited
directly from neural network.

This paper is also among the first social science studies that make
use of embeddings as part of empirical analyses. As another form of
unstructured data, text data have recently gained much attention in
economic analyses; see \citet{gentzkow2019} for a thorough review
of the machine learning applications. \citet{Kozlowski_2019} use
word embeddings to understand cultural norms. \citet{gentzkow2019measuring}
analyze political polarization using congressional speeches as text
data. \citet{hobergetal2016} use text data from firms' 10-K product
descriptions across industries to classify competing products and
construct a product location space as we do in our paper. Unlike their
paper, however, we use image embeddings and employ neural networks
as a classification method. We also use word embeddings as a way of
validating the image embeddings. Moreover, we focus on a particular
industry as opposed to multiple industries and utilize detailed structured
and unstructured data about product offerings. 

\subsubsection*{Merger and Product Differentiation\label{subsec:Merger-and-Product}}

Market structures and endogenous product differentiation have been
important themes in the field of economics. In a theoretical paper,
\citet{mazzeo2018} find that the effects of a merger on product differentiation
can be ambiguous, implying that the question is more of empirical
research, as in empirical industrial organization.\footnote{\citet{mankiw1986free} theoretically consider a more general question
of oligopolistic competition and product differentiation and again
suggest that the direction of entry bias can be unclear.} \citet{sweeting2013} studies the dynamic aspects of product differentiation
in the radio industry and finds some evidence suggesting that increased
concentration increases variety. \citet{fan2013} finds in newspaper
markets that mergers between local competitors has effects on vertical
differentiation, leading to a decrease in the news quality. In related
work, \citet{fan2020competition} consider multi-product firms in
smartphone markets and show how mergers may lead to a decline in the
number and variety of products, and \citet{wollmann2018trucks}'s
analysis of a commercial truck industry implies the opposite direction
of the merger effect on product variety. Apart from this important
line of work, the literature using a structural approach to study
the effects of merger on non-price and potentially unstructured attributes
is rather scarce due to difficulties in modeling endogenous product
offerings with high-dimensional attributes. This is in contrast to
merger effects on price, which have received relatively more attention
in the literature, e.g., building on the structural framework of \citet{nevo2001measuring}.

The structural approach has been complemented in the literature by
studies of merger and concentration from a viewpoint of treatment
effects and program evaluation. \citet{berry2001} and \citet{sweeting2010effects}
document the effects of mergers on product variety in local radio
markets.\footnote{See also \citet{atalay2020post} for a related merger analysis in
consumer goods markets. They find that mergers lead to dropping of
products that are dissimilar to existing ones, where the measure of
dissimilarity is calculated based on structured attributes.} \citet{hastings2004vertical} and \citet{ashenfelter2008effect}
study the effects of mergers on prices using the difference-in-differences
and instrumental variable methods, respectively. Although the treatment
effect approach to merger analyses has limitations (\citet{nevo2010taking}),
it is still suitable to highlight the rich information contained in
the visual dimension of product attributes previously neglected in
the literature on mergers.\footnote{See \citet{angrist2010credibility} and \citet{nevo2010taking} for
discussions on how the two approaches can complement each other and
what their pros and cons are.} We introduce unstructured data of images and related machine learning
methods to derive new insights into creative product differentiation
and its relationship to mergers. On the other hand, all the empirical
studies listed in this and the previous paragraphs use structured
data to construct response variables including price and non-price
attributes (e.g., product variety measures).\footnote{For example, \citet{fan2013} uses data on the number of opinion section
staff members, the number of reporters, local news ratio, variety,
frequency of publication, and edition. \citet{sweeting2013} uses
Neilson data on broadcasts. \citet{berry2001} use the number of stations
and programming formats as product offerings.} As mentioned, we cannot find in our analysis the merger effects on
the traditional product offerings measures. We believe this paper
is a good starting point for introducing embeddings into economic
analyses.

\subsubsection*{Machine Vision}

Recognizing letters (e.g., distinguishing handwritten ``G'' from
``Q'') is one of the most well-studied areas of machine vision as
is done with the MNIST database (\citet{lecun2010mnist}). Our paper,
however, is one of the first that applies machine vision techniques
to recognizing the \textit{style} of font images (e.g., distinguishing
typeface ``G'' from ``\textsf{G}''), which is a more challenging
vision problem. \citet{tenenbaum2000separating} propose bilinear
models that separate style and content with fonts as one of the examples.
\citet{o2014exploratory} develop a method for searching fonts using
relative attributes based on the work on attributes and whittle search
by \citet{parikh2011relative} and \citet{kovashka2012whittlesearch}.
\citet{campbell2014learning} develop a procedure for learning a font
manifold by parametrizing font shapes and reducing the dimension of
the resulting model.

The method of training the font embeddings builds on the works of
\citet{schroff2015facenet} and \citet{wang2014triplet}, who develop
a face recognition algorithm that directly learns an embedding for
images via neural network training. \citet{schroff2015facenet} show
that their approach performs substantially better than the earlier
approaches of training a classification network for face recognition,
such as those in \citet{taigman2014deepface} and \citet{sun2015deeply}.
The former approach is suitable for our purpose, as the procedure
produces embeddings as the intermediate output of the classification
algorithm. Although we are not directly interested in the classification
of font identity, embeddings serve as our object of primary interest.

Fonts can be viewed as fashion products. Our quantitative analysis
of the trend in font style is related to work by, e.g., \citet{al2017fashion},
\citet{mall2019geostyle}, and \citet{yu2019thinking}, who apply
advanced machine vision techniques they develop to recognize the style
of clothes and shoes in the fashion industry and understand the trend.
The analysis of the visual attributes of design products has also
been considered, e.g., in \citet{burnap2016estimating} and \citet{dosovitskiy2016learning}
using deep generative models with applications to furniture and automobile
designs. However, none of the studies conduct causal analyses to answer
economic questions.

\subsection{Organization of the Paper}

In the next section, we provide the background about the font industry
and the online marketplace for fonts considered in this paper. Section
\ref{sec:Data} describes the data obtained from this market. In Section
\ref{sec:Construction-of-Embeddings}, we construct the embedding
and the product space using the neural network. In this section, we
demonstrate that the embeddings are meaningful by calculating the
mutually shared information between the font embeddings and tags.
Section \ref{sec:Effects-of-Merger} presents the causal analysis
of a merger using the embeddings. Section \ref{sec:Conclusion} concludes.
Sections \ref{sec:Neural-Network-Training}--\ref{sec:Supplemental-Findings-for}
in the Appendix respectively present the internal evaluation of the
trained network, the descriptive analyses of style trends, and supplemental
results from the merger analysis.

\section{Online Marketplace for Fonts\label{sec:Backgrounds}}

We consider the world's largest online marketplace \href{https://www.myfonts.com}{MyFonts.com}
that sells around 30,000 different fonts. This market is a superset
of all major global online stores for fonts. MyFonts.com and the other
stores are all owned by Monotype Inc. A font is a delivery mechanism
for typefaces. Therefore, fonts are sold as a piece of software, for
which consumers purchase a license. Licenses are protected by the
End User License Agreement (EULA). Mostly two types of licenses are
sold to consumers. A web font license allows fonts to be displayed
on a website, and a desktop license is for printed material. The marketplace
also serves as a platform for third-party fonts. As a result, it sells
fonts designed by foundries owned by Monotype as well as fonts from
third-parties foundries.\footnote{A foundry is a group of designers who create fonts.}

\begin{figure}
\noindent \begin{centering}
\includegraphics[scale=0.11]{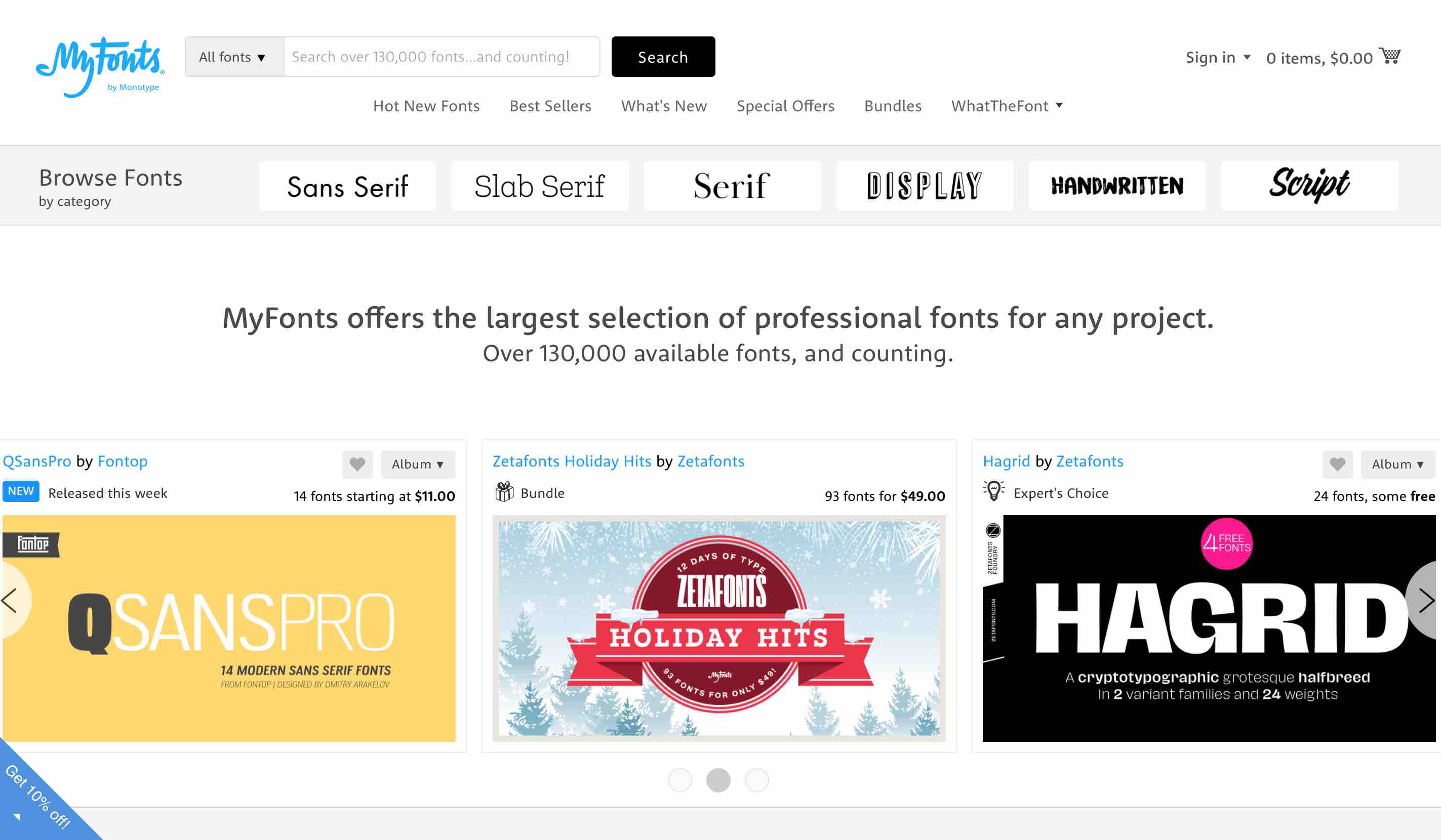}
\par\end{centering}
\caption{Home Page of MyFonts.com}
\label{fig:mainpage}
\end{figure}

\begin{figure}
\noindent \begin{centering}
\includegraphics[scale=0.15]{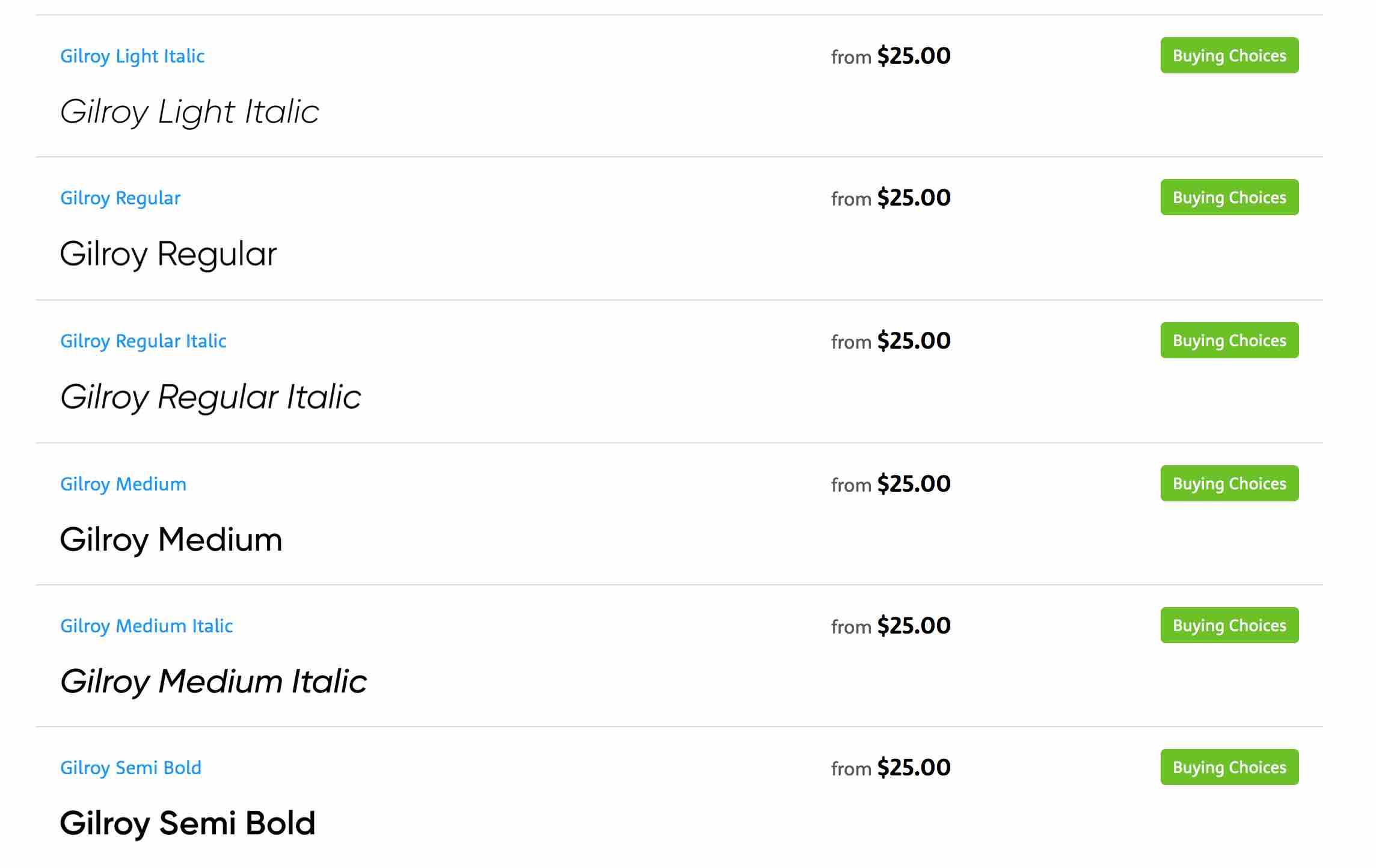}
\par\end{centering}
\caption{A Font Family Page on MyFonts.com}
\label{fig:familypage}
\end{figure}

Figure \ref{fig:mainpage} shows the home page of MyFonts.com.
 An example of a font family page on MyFonts.com is captured in Figure
\ref{fig:familypage}.
 In the font industry, a \textit{family} represents the identity of
a font, which name is the name of the font. A font family is consist
of several different \textit{styles}, such as regular, light, bold,
and italic. Figure \ref{fig:familypage} shows different styles of
Gilroy family. In this market, typical consumers are independent designers
who use fonts as intermediate goods. They produce printed material
(e.g., posters, pamphlets, cards), for which a desktop license is
purchased, or webpages and digital ads, for which a web license and
digital ads license are purchased, respectively. In the data collection
period of 2012--2018, around 2,400,000 purchases were made.

\section{Data\label{sec:Data}}

\subsection{Overview}

Our sample comprises data from 2002 to 2017. The dataset includes,
in total, 28,659 fonts and 2,446,604 orders. The main information
contained in the dataset is product attributes and transactions for
each consumer. The unstructured high-dimensional attributes include
images of typefaces and tags (i.e., descriptive words assigned by
producers or consumers). The structured attributes include price,
category types, license types, the number of languages supported,
the number of glyphs supported, the foundry and designer information,
and the date of introduction in the market. There are roughly six
category types: sans serif, serif, slab serif, display, handwritten,
and script. Transaction data include information on individual orders
made by consumers and consumer characteristics such as the country
and city of origin. We will revisit this dataset in Section \ref{sec:Effects-of-Merger}
for the merger analysis. For now, we focus on the unstructured data.

\subsection{Visual Attributes}

\begin{figure}
\begin{centering}
$\qquad\qquad\qquad$\includegraphics[scale=0.2]{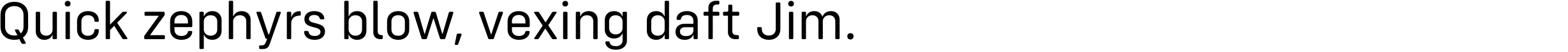}
\par\end{centering}
\begin{centering}
\vspace{0.4cm}
$\qquad\qquad\qquad$\includegraphics[scale=0.2]{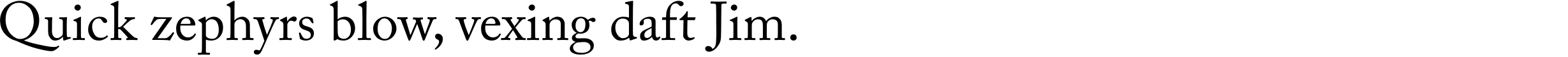}
\par\end{centering}
\begin{centering}
\vspace{0.4cm}
$\qquad\qquad\qquad$\includegraphics[scale=0.2]{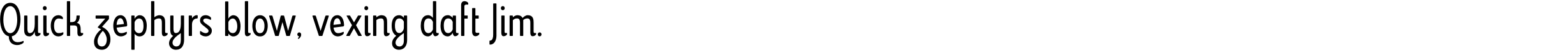}
\par\end{centering}
\begin{centering}
\vspace{0.4cm}
$\qquad\qquad\qquad$\includegraphics[scale=0.2]{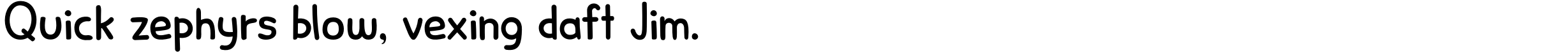}
\par\end{centering}
\begin{centering}
\vspace{0.4cm}
$\qquad\qquad\qquad$\includegraphics[scale=0.2]{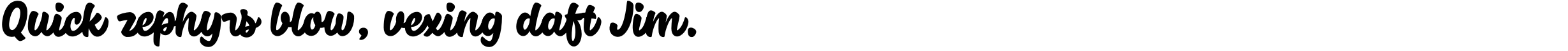}
\par\end{centering}
\begin{centering}
\vspace{0.4cm}
$\qquad\qquad\qquad$\includegraphics[scale=0.2]{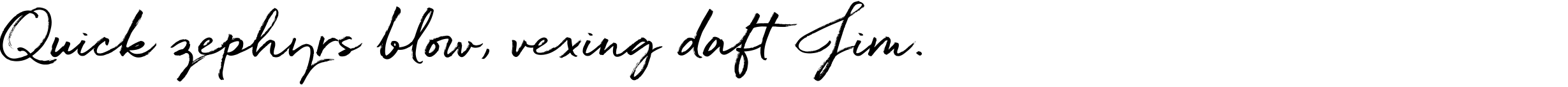}
\par\end{centering}
\caption{Examples of a Pangram (by category types)}
\label{fig:pangrams}
\end{figure}

Fonts are displayed on the webpage using pangrams that contain all
the alphabet letters in one sentence.\footnote{In font markets, many different pangrams are used: ``The quick brown
fox jumps over a lazy dog.'' ``Six quite crazy kings vowed to abolish
my pitiful jousts.'' ``Quincy Jones vowed to fix the bleak jazz
program.'' ``Mozart's jawing quickly vexed a fat bishop.'' Here
we chose to use one of the shortest pangrams to minimize the size
of the image.} Pangrams effectively capture important design elements that cannot
be seen in individual letters, such as spacing, deep-height, up-height,
and ligature. We use pangrams as direct inputs in the neural network
in order to mimic a consumer's actual perception of the products.
Figure \ref{fig:pangrams} shows the examples of a pangram that roughly
correspond to the product categories (i.e., sans serif, serif, slab
serif, display, handwritten, and script).
 The format of pangram images is a bitmap with $200\times1000$ pixels,
where each pixel is a greyscale with a value between 0 and 255. We
use random crops of $100\times100$ pixels as inputs in the network.

\section{Construction of Embeddings\label{sec:Construction-of-Embeddings}}

\subsection{Neural Network Embedding}

We employ a method in which the network directly learns a mapping
from pangram images to a compact Euclidean space. This mapping is
called an embedding. We map each pangram to a 128-dimensional embedding,
denoted as $f(x)\in\mathbb{R}^{d}$ for pangram image $x$ with $d=128$.\footnote{It is important to have sufficient dimensions so that the neural network
can allow for variation in the embedding space based on the actual
images. The embedding with a larger number of dimensions would perform
better, but it requires more training data to achieve the same level
of accuracy while avoiding the risk of overfitting. We additionally
normalize each embedding so that it lies on a 128-dimensional hypersphere,
i.e., $\left\Vert f\right\Vert _{2}=1$ with the Euclidean norm $\left\Vert \cdot\right\Vert _{2}$.} The Euclidean distance in the resulting embedding space corresponds
to the measure of similarity of font shape. For training the network
embedding, we adapt a modern algorithm developed by \citet{schroff2015facenet}
for face recognition. Their algorithm determines the identity of a
person based on face images in two steps. In the first step, a deep
convolutional network is trained to learn an embedding space of faces.
The rationale is that similar faces should occur closer in the embedding
space than dissimilar faces. In the second step, the identities of
the images are classified by choosing a threshold in the space below
which the embeddings have the same identity. They show that this approach
performs substantially better than the earlier approaches of training
a classification network (\citet{taigman2014deepface}, \citet{sun2015deeply}).
Our algorithm builds on \citet{schroff2015facenet}, which approach
is suitable for our purpose. First, pangram images can be classified
based on a structure similar to that for face images. In the dataset
of face images, multiple images are associated with the same identity.
Analogously, multiple styles are associated with the same font family
as detailed below. Second, the approach produces embeddings as the
intermediate output of the algorithm. Although we are \textit{not}
directly interested in the classification of font identity, embeddings
serve as our object of primary interest.

\subsection{Triplet Loss and Network Training}

The neural network is trained to produce embeddings and classify fonts
that are within the same family in the resulting embedding space.
In practice, we accomplish this by constructing triplets of images.
Triplet $i$ comprises anchor $x_{i}^{a}$, positive $x_{i}^{p}$,
and negative $x_{i}^{n}$. An anchor is a pangram image of a given
font family (e.g., Helvetica), positives are pangram images of the
same family but different styles (e.g., Helvetica Regular, Helvetica
Light, Helvetica Bold, Helvetica Italic), and negatives are pangram
images of different families (e.g., Time New Roman).\footnote{To be precise, pangram images here refer to crops of the images. Therefore,
we also use different crops within the same pangram as positives.} The way triplets are sampled is analogous to that in the face recognition
problem, where positives are images of the same person as the anchor
and negatives are images of different persons. For triplet $(x_{i}^{a},x_{i}^{p},x_{i}^{n})$
in the entire set of font images, we enforce the following inequality
during the network training:
\begin{align}
\left\Vert f(x_{i}^{a})-f(x_{i}^{p})\right\Vert _{2}^{2}+\alpha & \leq\left\Vert f(x_{i}^{a})-f(x_{i}^{n})\right\Vert _{2}^{2},\label{eq:triplet}
\end{align}
where $f(x)\in\mathbb{R}^{128}$ is the embedding of image $x$, $\left\Vert \cdot\right\Vert _{2}$
is the Euclidean norm, and $\alpha$ is an enforced margin. That is,
we ensure that the distance between an anchor and positive is smaller
than that between the anchor and a negative.\footnote{The choice of triplets is very important for the fast convergence
of the algorithm. As such, we make sure we choose sufficiently many
triplets with hard positives and negatives, namely triplets that violate
\eqref{eq:triplet}.} The margin $\alpha$ allows the images for one font family to stay
closer in the embedding space, while still discriminating the images
of other families.

Then, a triplet-based loss function that is minimized in the network
training is
\begin{equation}
L=\sum_{i}^{N}[\left\Vert f(x_{i}^{a})-f(x_{i}^{p})\right\Vert _{2}^{2}-\left\Vert f(x_{i}^{a})-f(x_{i}^{n})\right\Vert _{2}^{2}+\alpha]_{+}.\label{eq:loss}
\end{equation}
We optimize this objective using stochastic gradient descent (SGD;
\citet{bottou2010large}). SGD is an iterative method for optimizing
an objective function---in our case, for creating a reasonable embedding
space. Because our dataset size is in gigabytes, it would be computationally
challenging to compute the gradient of the entire dataset and optimize
it using a more traditional optimization algorithm (e.g., Nelder-Mead
or the conjugate gradient method). SGD can be regarded as a stochastic
approximation of gradient descent optimization. It replaces the actual
gradient by an estimate of the gradient.

We use approximately 20,000 images of fonts to train the neural network.
The training iteratively improves the parameters of the network using
small batches of images to estimate the gradient and then update the
parameters accordingly. As the gradient is evaluated at more batches,
the parameters in the network are adjusted. The training of the network
is completed when the loss function reaches below a certain threshold.
Section \ref{subsec:Details-of-Network} in the Appendix contains
the details.

As internal evaluation, we show in Section \ref{sec:Internal-Evaluation}
how the neural network performs in the original classification task
of identifying the font family. Although we are interested in the
embeddings and not the classification, it is important to evaluate
the embeddings based on their ability to classify. If the embeddings
perform poorly, then they are not likely to generate a reliable embedding
space. As mentioned, the classification of font families is conducted
by thresholding the distances between embeddings. Overall, the neural
network embeddings perform well in differentiating between fonts in
different families.

\subsection{Constructed Product Space}

\begin{figure}
\noindent \begin{centering}
\includegraphics[scale=0.15]{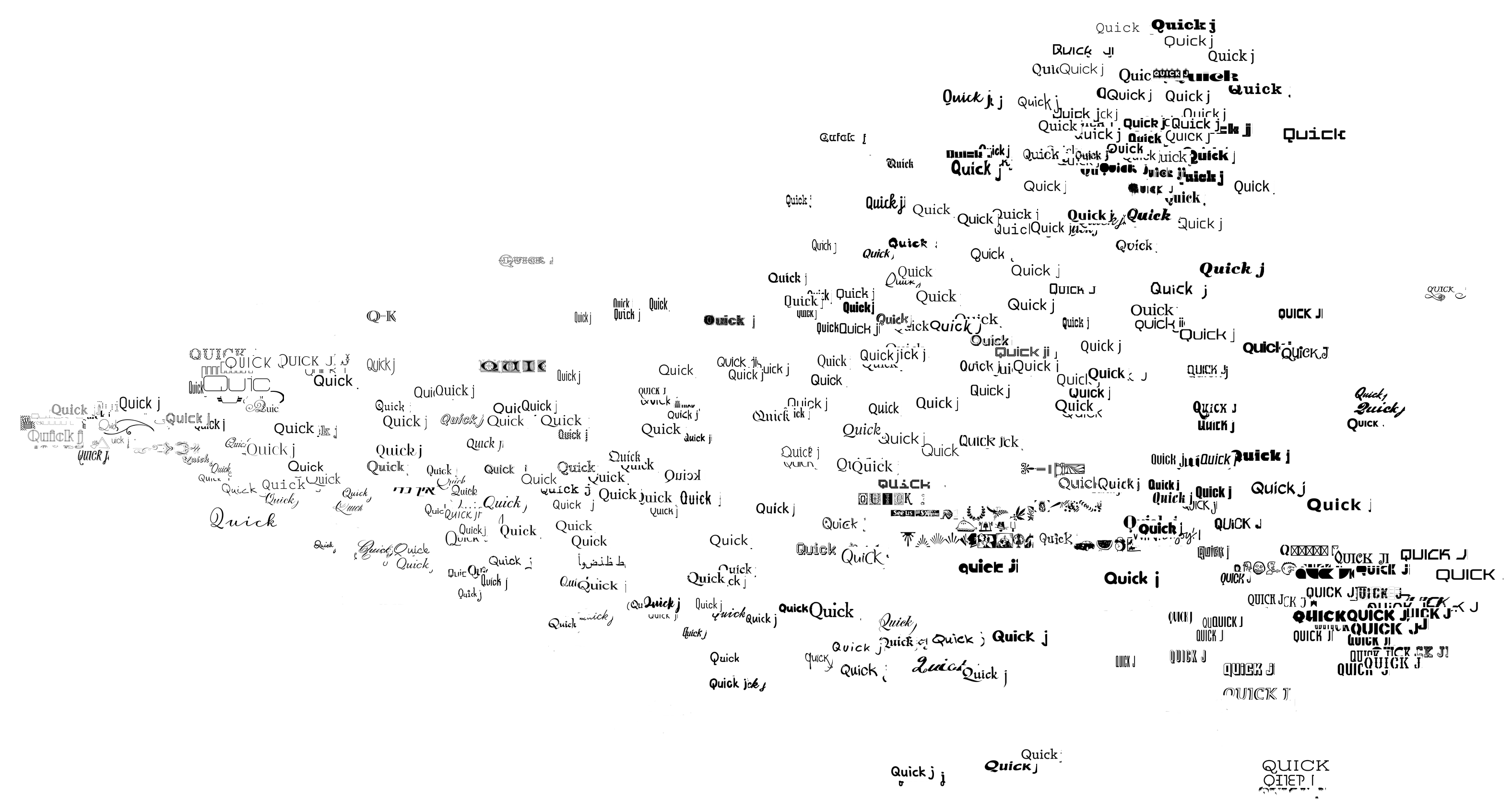}
\par\end{centering}
\caption{Two-Dimensional Visualization of the Space for Fonts}
\bigskip{}

{\footnotesize{}}%
\noindent\begin{minipage}[t]{1\columnwidth}%
{\footnotesize{}Note: This figure is created using a randomly selected
subsample of 400 fonts projected onto two dimensions using t-SNE.}%
\end{minipage}{\footnotesize\par}

\label{fig:prod_sp}
\end{figure}

The trained neural network produces embeddings with 128 dimensions,
which define a 128-dimensional space of font products. Figure \ref{fig:prod_sp}
visualizes this space by projecting it onto a two-dimensional space
using t-distributed stochastic neighbor embedding (t-SNE).\footnote{t-SNE is a useful tool to visualize high-dimensional data in a two
or three dimensional space. In our setting, we visualize 128-dimensional
objects in a two-dimensional space.} For expositional purposes, the thumbnail is created with the word
``Quick'' that is cropped from the full pangram. Each thumbnail
corresponds to the embedding of the \textit{regular style} of each
font family as a representative style. Therefore, for example, boldface
in this figure is a design aspect of a particular font family, \textit{not}
a bold style within a family. A visual inspection reveals that different
styles of fonts are well-clustered together even in this two-dimensional
projection. More specifically, we can see that the western region
is populated with fonts that have narrower characters and the eastern
region with fonts that are thicker. In addition, the shape of fonts
in the northeast is more geometric while the shape in the southwest
is more curly. It is worth noting that these patterns are observed
even in this space with restricted dimensions. The space we base our
economic analysis on is the original 128-dimensional space.

\begin{figure}
\noindent \begin{centering}
\includegraphics[scale=0.6]{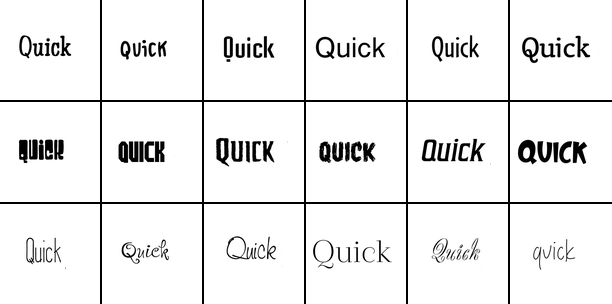}
\par\end{centering}
\caption{Nearest Neighbors (Collected in Each Row) in the 128-Dimensional Product
Space}

\label{fig:closest_fonts}
\end{figure}

To further understand how well the fonts are clustered in the 128-dimensional
product space, we present in Figure \ref{fig:closest_fonts} the examples
of regular fonts that are close to each other in the space. In Figure
\ref{fig:closest_fonts}, the images of six nearest neighbors in terms
of the Euclidean distance are listed in each row of the table.
 Even though we only list regular fonts, it is clear that fonts with
different thicknesses are clustered together in the space. Also, the
middle and the last rows show that geometric and curly fonts tend
to cluster together, respectively.

Unlike individual embeddings, the distance metric endowed in this
space has the clear interpretation of visual similarity and is more
robust to model specifications. These aspects allow us to use the
distance metric as the main building block in the economic analysis
in Section \ref{sec:Effects-of-Merger}.

\subsection{Relevance of Embeddings}

Although the neural network embedding performs well in terms of classifying
fonts of similar shapes, good predictive performance does not necessarily
imply the relevance of the embeddings for economic analyses. To address
this concern, we verify that the visual attributes captured in the
resulting embeddings are relevant to economic agents' perceptions
by measuring (a generalized notion of) their correlation with ``perceived''
attributes. For the latter, we use information from \textit{tags},
which are short descriptive words assigned to each font family by
font designers and consumers. Examples of tags are ``curly,'' ``flowing,''
``geometric,'' ``organic,'' ``decorative,'' and ``contrast.''
These descriptive words of fonts are also high-dimensional, as the
tags include nearly 30,000 different words. Therefore, we consolidate
tags that are synonyms to create meta-tags that encompass many related
words (e.g., ``handwritten'' and ``cursive'' would be in the same
meta-tag). To this end, we create clusters of tags using a standard
word embedding ``Word2vec'' (two-layer neural network) by \citet{mikolov2013efficient}.

To measure the relevance between the image and word embeddings, we
create clusters of the image and word embeddings, respectively, using
$K$-means clustering.\footnote{To perform $K$-means clustering of font shape, we use the obtained
128-dimensional image embeddings directly clustered into 60 clusters
using the elbow method. To perform $K$-means clustering of tags,
we use 100-dimensional word embeddings that are first reduced to 10-dimensional
embeddings and then clustered into 6 clusters using the elbow method.
} Then, we measure how well the clusters of word embeddings match the
clusters of image embeddings by using mutual information. Specifically,
we use the normalized mutual information (NMI).
\begin{align*}
NMI(F,W) & =\frac{I(F,W)}{\{H(F)+H(W)\}/2}
\end{align*}
In this formula, $F$ is the distribution of clusters based on font
image embeddings, $W$ is the distribution of clusters based on word
embeddings, $H(\cdot)$ is information entropy, and $I(F,W)=H(F)-H(F|W)$
is the mutual information between $F$ and $W$. The NMI can be interpreted
as how informative $W$ is in determining $F$. Its value ranges between
$0$ and $1$ (the value $0$ implies that $W$ contains no information
regarding $F$). 
\begin{table}
\begin{centering}
\begin{tabular}{ccc}
\toprule 
$F$ & $W$ & $NMI(F,W)$\tabularnewline
\midrule 
Image Embeddings & Word Embeddings & 0.473\tabularnewline
Industry Categories & Word Embeddings & 0.261\tabularnewline
\bottomrule
\end{tabular}
\par\end{centering}
\caption{Normalized Mutual Information between Image Clusters and Word Clusters
(first row), Compared to Baseline (second row)}
\label{tab:NMI}
\end{table}
Table \ref{tab:NMI} reports the values of NMI for two different pairs
of distributions. The first row corresponds to the NMI for the image
and word clusters described above. We obtain $NMI(F,W)=0.473$, which
is quite promising. To compare this value with the baseline case,
the second row of the table shows the NMI when the pre-existing product
categories are used as structured attributes instead of the image
embeddings. In general, when images were treated as unobserved product
attributes, then the main structured attribute available in many design
products is product categories. Recall that sans serif, serif, slab
serif, display, handwritten, and script are such product categories
defined by the font industry.\footnote{Product categories are generally available to analysts, but it is
specific to this online marketplace that tags are observed. This motivates
the setup of having tags as the ground truth in this section.} In this case, we obtain $NMI(F,W)=0.261$, which is roughly only
half of the value in the first case. These results suggest that the
learned image embeddings arguably capture economic agents' perceptions
better than the structured attributes and can be relevant for economic
analyses.

\section{Effects of Merger on Design Differentiation\label{sec:Effects-of-Merger}}

\subsection{Background}

On June 15, 2014, one of the major font foundries called FontFont
was acquired by Monotype. At the time, Monotype sold fonts created
by the foundries it owned as well as by third-party foundries. Before
the merger, FontFont had sold its fonts through MyFonts.com as a third
party. We study the causal effect of this merger on the change in
the product differentiation decisions of the merging firm (i.e., FontFont
foundry). As we show below, a favorable aspect of this market for
merger analysis is that price seems to play little role in competition
so that we can focus on product differentiation as the main response
variable to merger. The major channel for product differentiation
is the design of font shapes. This creative decision of a foundry
may be affected by the merger through many different channels. By
merging, the firms could increase efficiency by reducing transaction
costs. If costs are reduced, the merged firm might find it profitable
to create more experimental products. The merged firm may also be
concerned with cannibalization---that is, competition among their
own products---which would increase the diversity of product design.
On the other hand, if Monotype has a preemptive motive with the merger,
it will crowd its products to prevent entries. In the presence of
these opposing factors, whether the merger increases the design diversity
is an empirical question. We answer this question using the embeddings
we created as the main ingredient of our statistical model.

\subsection{Design Differentiation Measures}

In this merger analysis, the outcome of interest is the degree of
design differentiation. We construct two measures for design differentiation
based on the constructed embeddings. The first measure is the \textit{distance
to Averia}. We calculate the Euclidean distance between an individual
font and a benchmark font:\footnote{For each font embedding in our economic analyses, we use a representative
embedding in a given family, namely the embedding of a regular style.} for image $x_{i}$ of font $i$,
\begin{align*}
D_{i}^{A} & =\left\Vert f(x_{i})-f_{averia}\right\Vert _{2},
\end{align*}
where $f(\cdot)\in\mathbb{R}^{128}$ is the embedding and $f_{averia}$
is the embedding of a benchmark font called ``Averia,'' which is
calculated by averaging the values of the embeddings of all the existing
fonts in the market. The distance measure $D_{i}^{A}$ is intended
to capture the degree of product differentiation of font $x_{i}$. 

In addition to $D_{i}^{A}$, we consider the following \textit{gravity
measure}: for image $x_{i}$ of font $i$,
\begin{align*}
D_{i}^{G} & =-\sum_{j\neq i}\frac{1}{\left\Vert f(x_{i})-f(x_{j})\right\Vert _{2}},
\end{align*}
where the sum is for all other fonts $j$'s in the market. This measure
effectively captures how font $i$ is located relative to other fonts
$j$'s in the product space; it takes a large value when $i$ is individually
far from all $j$'s and a small value when it is close to any of them.
Compared to the distance to Averia, the gravity measure acknowledges
the aspect of spatial competition among designers. For illustration,
consider the interval $[0,1]$ as the product space. Suppose two existing
fonts are located at $\{0,1\}$ in terms of their shapes, and the
third font chooses to enter a location between one of $\{0,1/2,1\}$.
The shape of the third font $i$ would be most differentiated in terms
of $D_{i}^{G}$ if it is located at $\{1/2\}$. On the other hand,
$i$ would be the most differentiated product in terms of $D_{i}^{A}$
if it is located at one of $\{0,1\}$ (even though these points are
already populated).

Finally, we aggregate each measure for all fonts created by foundry
$k$ in period $t$. In particular, for $j\in\{A,G\}$, we construct
\begin{align*}
\bar{D}_{kt}^{j} & =\frac{1}{\left|I_{kt}\right|}\sum_{i\in I_{kt}}D_{i}^{j},
\end{align*}
where $I_{kt}$ is the set of all fonts created by foundry $k$ in
period $t$. In the subsequent merger analysis, we consider both $\bar{D}_{kt}^{A}$
and $\bar{D}_{kt}^{G}$ as the outcome variables, henceforth referred
to as the \textit{mean deviation }and\textit{ gravity measures}, respectively.
Despite the distinct feature, both $\bar{D}_{kt}^{A}$ and $\bar{D}_{kt}^{G}$
are meant to succinctly capture each foundry's creative decision of
product differentiation in terms of font design every period. We show
that the subsequent empirical analysis is robust to the choice of
the measure between the two.

\subsection{A Theoretical Example for Merger and Differentiation}

We first consider a simple Hotelling-style model to illustrate that,
in the market for fonts, the degree of product differentiation may
be larger with merger than without while price may remain the same
and consequently welfare may be greater with merger. The prediction
from this theoretical model serves as the motivation of the subsequent
empirical analysis. Consider two representative foundries each of
which produces one font by locating its shape in a unit interval $[0,1]$.
Foundry 1 chooses location $a$ and foundry 2 chooses location $1-b$.
Based on tastes for fonts, consumers are located at $x$ distributed
uniformly over $[0,1]$. Each consumer experiences aesthetic ``transportation
cost'' $t$, which is incurred by ``traveling'' from her specific
taste to an available font on the market. Consumers then pay $p_{1}$
for foundry 1's font and $p_{2}$ for foundry 2's font. The utility
of consumer located at $x$ from buying font 1 is
\[
V_{1}(x)=u-p_{1}-t|a-x|,
\]
where $u$ is a utility parameter, and the utility from buying font
2 is
\[
V_{2}(x)=u-p_{2}-t|1-b-x|.
\]
Finally, consumers' utility if no product is purchased is normalized
to be $V_{0}=0$. This model is slightly more general than what is
considered in \citet{berry2001}. 

When travel cost $t$ is high relative to $u$, the firms in the Hotelling
model may not compete with each other (\citet{economides1989}). In
this case, the model predicts (as shown below) that some consumers
may not buy the product (i.e., $V_{0}\geq V_{1}(x)$ and $V_{0}\geq V_{2}(x)$
for some $x$) and each firm becomes a \emph{local monopoly}, only
selling to consumers with positive net surplus. We derive the equilibrium
location and price under local monopoly. We start by finding foundry
1's profit function, which depends on the share of consumers who buy
its product. To find this share, consider the consumer who is indifferent
between buying and not buying font 1: $V_{1}(x)=0=V_{0}$. Let $x_{1}$
be the westmost location of a consumer who will buy font 1. Solving
for $x_{1}<a$, we have $x_{1}=a-(u-p_{1})/t$. Let $x_{2}$ be the
eastmost location of a consumer who will buy font 1. Solving for $x_{2}>a$,
we have $x_{2}=a+(u-p_{1})/t$. Assume that the marginal cost is zero,
which is plausible in this market. Then foundry 1's \textit{operating}
profit function of choosing $a$ and $p_{1}$ would be
\[
\pi_{1}(a,p_{1})=p_{1}(a-x_{1})+p_{1}(x_{2}-a)=2p_{1}(u-p_{1})/t.
\]
Maximizing this profit yields the optimal price of $p_{1}=u/2$. Note
that location $a$ and the price $p_{2}$ chosen by foundry 2 are
not part of foundry 1's profit function (and analogously for foundry
2's profit), which reflects the fact that each firm is a local monopoly.
As a result, there are multiple possible equilibrium locations for
$a$ (and for $b$), subject to the fact that the share of consumers
who buy font 1 cannot overlap with the share who want to buy font
2. We can analogously derive the optimal price for foundry 2, which
yields $p_{2}=u/2$, the same as the optimal $p_{1}$.

Consider a concrete example of the model by assume $t=1$ and $u=1/3$.
Further assume there is an fixed entry cost $F=1/20$. Then, the optimal
prices are $p_{1}=p_{2}=1/6$ and $(1/3,2/3)$ is one of the equilibrium
location. Under this solution, the consumer at the center of the
space will not buy anything because $V_{1}(1/2)=V_{2}(1/2)=-1/3<V_{0}=0$.
Therefore, the location and price are the equilibrium under local
monopoly. However, this is an inefficient equilibrium as only $2/3$
of consumers buy the fonts; consumers $[1/6,1/2]$ buying font 1 and
$[1/2,5/6]$ buying font 2. Still, neither firm has an incentive to
deviate and introduce an new product that yields enough profit justifying
the fixed cost.

Now consider a market where both foundries are merged under the same
parameter values for $(t,u,F)$. The merged firm may introduce three
products in the location $(1/6,1/2,5/6)$. To see why this is a local
monopoly equilibrium, note that none of the products compete with
each other; consumers from $[0,1/3]$ buy font 1, consumers from $[1/3,2/3]$
buy font 2, and consumers from $[2/3,1]$ buy font 3. Moreover, this
is a unique and efficient equilibrium as the products serve all the
consumers in the market. In terms of product differentiation, note
that the maximum differentiation (i.e., the distance between furthest
endpoints) is $2/3$, larger than $1/3$ without the merger. Similarly,
in terms of our gravity measure, differentiation is approximately
$-2.0$, larger than $-2.19$ without the merger. To see why the price
remains the same, the merged firm's joint profit function involving
all 3 products is $\pi_{1}(p_{1})+\pi_{2}(p_{2})+\pi_{3}(p_{3})$
where $\pi_{1}$, the profit for font 1, does not depend on the prices
of fonts 2 and 3 and so on. Then, the optimal price would be $p_{1}=p_{2}=p_{3}=1/6$,
the same as the optimal price without the merger. In this case, the
profit is $1/6$ with the three products, compared to $1/9$ which
is the aggregate profit of the two firms without the merger. Because
more consumers are served with the merger while price remains the
same, the welfare would be also greater with the merger than without.

For all other two-product equilibria without the merger that are close
to $(1/3,2/3)$,\footnote{Specifically, they are equilibria where (approximately) $a\in[0.3157,1/3]$
and $b\in[2/3,1-0.3157]$.} a similar pattern is predicted: increased differentiation with the
merger while price stays the same. When the locations move further
away from $(1/3,2/3)$ toward $(1/6,5/6)$, then one of the firms
may introduce another product, which results in a three-product equilibrium.
An example would be $(1/6,1/2,5/6)$ and $p_{1}=p_{2}=p_{3}=1/6$,
which are identical to the unique equilibrium location and price with
the merger. For all other three-product equilibria without the merger,\footnote{They are equilibria where (approximately) $a\in[1/6,0.3157]$, $b\in[1-0.3157,5/6]$,
and the third product locating at a point in $[0.3882,1/2]$ by one
of the firms and the exact locations are dependent to one another.} differentiation increases with the merger and price may stay the
same (for the single-product foundry) or decrease (for the two-product
foundry). In all these other cases, since the price is no larger with
the merger while the market is fully served, the welfare would be
no smaller than without the merger.

To summarize the prediction under local monopoly, for a large range
of equilibria, we expect increased differentiation and invariant price
with merger than without. For all equilibria, there is a welfare gain
with merger. The model's prediction is starkly different when there
is no local monopoly. This occurs when $t<u$. In contrast to the
prediction under local monopoly, differentiation falls and price increases
with merger than without. Therefore, in this case, the welfare consequence
is ambiguous.

Given the theoretical findings, the effects of merger on product differentiation
and price (and consequently on welfare) in the market for the current
design products remain an empirical question. Although this illustration
is based on a simple stylized model, a similar intuition would continue
to hold when there are more than two foundries and each foundry produces
multiple fonts. Below, as our main focus of investigation, we empirically
show that the responses of product differentiation and price in the
real-world merger case of FontFont are in fact consistent with the
prediction of the model under local monopoly.

\subsection{Data and Empirical Strategy\label{subsec:Data-and-Empirical}}

For our empirical analysis, we construct the panel based on the dataset
described in Section \ref{sec:Data}. The cross-sectional units of
this panel are foundries and the time dimension is a bi-annual time
series between 2002 and 2017. Table \ref{tab:sum_stats} shows the
summary statistics for the variables (except the raw embeddings) in
the panel. All the variables are constructed to be foundry-level.
The mean deviation and gravity measures are the main response variables
in the merger analysis. Glyph count is an important product specification
and is considered to be related to quality.\footnote{Glyph count is the number of characters including special characters
in each font family. We calculate the average glyph count for all
fonts produced by each foundry each period. } The release frequency (i.e., maximum period between two releases),
sales, number of orders, average price, and age of foundries introduced
are the other control variables we use.
\begin{table}
\begin{centering}
\begin{tabular}{ccccc}
\toprule 
 & Mean & S.D. & Min & Max\tabularnewline
\midrule 
Mean Deviation & 0.41 & 0.11 & 0.18 & 0.74\tabularnewline
Gravity & -9.68 & 0.16 & -9.97 & -9.24\tabularnewline
Glyph Count & 419.03 & 775.58 & 32 & 9,844\tabularnewline
Release Frequency & 6.25  & 7.02  & 0.00  & 30.00 \tabularnewline
Sales (\$1K) & 601  & 1,886  & 0 & 24,636 \tabularnewline
Order Count & 7,517  & 23,239  & 1 & 360,435 \tabularnewline
Price per Order (\$) & 92.24  & 94.62  & 1.60  & 678.07 \tabularnewline
Age (Half Year) & 10.32  & 5.34  & 0.00  & 17.50 \tabularnewline
\midrule 
$N\times T$ & \multicolumn{4}{c}{$51\times32=1,632$}\tabularnewline
\bottomrule
\end{tabular}
\par\end{centering}
\caption{Summary Statistics (embeddings variables omitted)}
\label{tab:sum_stats}
\end{table}

Section \ref{sec:Trend-Analysis} in the Appendix contains a simple
descriptive analysis of the supply- and demand-side trends for the
shapes of fonts captured in the embeddings. The supply-side trend
plots the differentiation measures constructed in the previous subsection
for fonts newly introduced in the market every period. For the demand-side
trend, we construct and plot analogous measures for fonts purchased
every period. We find that, on average, newer fonts tend to be more
differentiated than older fonts. How such differentiation decisions
are affected by the change in market structure is the subject of the
merger analysis. We also find that the demand-side trend is markedly
stable, especially around the time of merger.\footnote{In general, the stable demand-side trend is consistent with the industry
norm. According to the industry experts we interviewed, font markets
tend not to experience seasonality or short-term trends in consumer
preferences, unlike in other design industries such as clothing. } This supports the argument that the change in the producer behavior
we find below in the merger analysis is not demand-driven.

To estimate the causal effect of merger on product differentiation,
we find a comparable group that would behave similarly as the treated
foundry (i.e., FontFont) if it were not for the merger. There are
challenges in using this approach. First, it is difficult to find
a single untreated foundry that resembles the treated foundry in terms
of the product differentiation measure (i.e., the mean deviation or
gravity measure). A naive average within a control group would be
a poor candidate. Second, candidate foundries for the control group
whose products are too similar to the treated foundry's products may
be in direct competition with the treated, which can create strategic
spillovers. To overcome these challenges, we use the synthetic control
method (\citet{abadie2003economic}, \citet{abadie2010synthetic})
and the proposed differentiation measures. First, the synthetic control
method compares the treated unit with a ``synthesized control unit''
obtained from a weighted average of the control group. Here, the weights
are estimated by minimizing the distance between the observed characteristics
including the outcome (as presented in Table \ref{tab:sum_stats})
of the treated unit and the weighted average of the characteristics
of the control group. For the control group that serves as the basis
for constructing a synthetic control, we use foundries whose merger
status has not changed during the study period. Second, although we
achieve a certain level of similarity in the differentiation measure
between the resulting synthetic control and the treated foundry, this
does not necessarily imply that they are competitors producing similar
products. This is because the value of the measure of each control
unit can be significantly different from that of the treated unit,
even if the weighted average is close to the latter. In fact, this
appears to be the case in our data as shown in Figures \ref{fig:synth-gravity}
and \ref{fig:synth-mean_dev} below. Therefore, we assume that strategic
spillovers between the treated unit and the control units are not
substantial.\footnote{See \citet{abadie2021using} for discussions on similar approaches
to minimize strategic spillovers between treated and control units.} Nonetheless, the advantage of our setting is that the embeddings
contain rich information about the design attributes of the fonts
created by the foundries. Therefore, we can reliably create a comparable
synthetic unit from the control group without their products being
necessarily visually similar to those of the treated foundry. In total,
our data include one treated foundry (FontFont) and 50 control foundries.

\subsection{The Effects of Merger}

\begin{figure}
\begin{centering}
\includegraphics[scale=0.45]{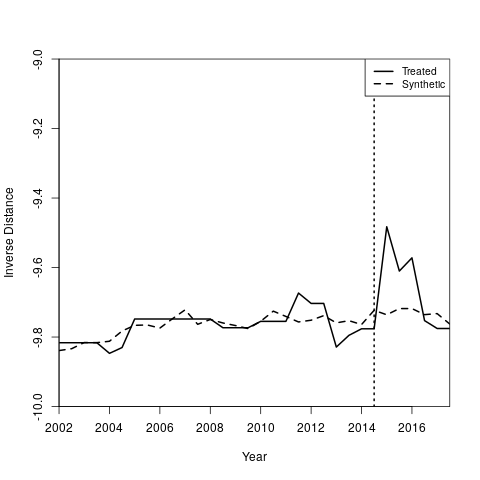}\includegraphics[scale=0.45]{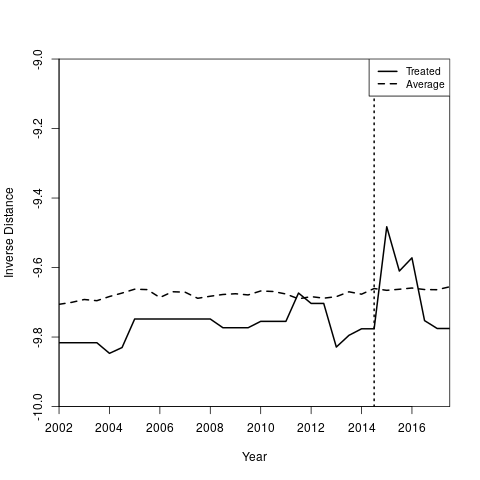}
\par\end{centering}
\caption{Trends of FontFont vs. Synthetic FontFont (left) and Naive Control
Group (right)---Using Gravity Measure}
\medskip{}

{\footnotesize{}}%
\noindent\begin{minipage}[t]{1\columnwidth}%
{\footnotesize{}Note: The solid line depicts the trend of FontFont
and the dashed line depicts the trend of the synthetic control (left)
or the naive average trend among all the control units (right). The
vertical dotted line in each figure indicates the time of the merger.}%
\end{minipage}\label{fig:synth}
\end{figure}
Figure \ref{fig:synth} captures the main results of our causal analysis
of merger using the gravity measure.\footnote{To reduce the scale, we take the logarithm of the positive part of
$\bar{D}_{kt}^{G}$ and then put back the minus sign.} The results of the analysis using the mean deviation measure are
presented in the Appendix. The solid line in the left panel presents
the trend of the gravity measure of the fonts \textit{newly} designed
by FontFont in a given period. Around the period where FontFont was
merged, which is indicated by a vertical line, the shape of FontFont's
fonts substantially differs from that of other fonts in the market.
This before and after comparison alone cannot yield the causal effect
of the merger, as other market conditions may have changed around
this period. Comparing this trend with the trend of the synthetic
FontFont, depicted by a dashed line, removes possible confounding
factors. First, the two trends before the merger appear to be close
to each other by construction. After the merger, however, FontFont
tends to produce more experimental fonts (i.e., fonts that are far
from others) relative to the trend of the synthetic control. We also
confirm that even if we backdate the period of acquisition (i.e.,
does not use the information of the timing of merger), the relative
increase still occurs around the same period as the vertical line.

To understand the virtue of our synthetic control method, we contrast
the left panel with the right panel in Figure \ref{fig:synth}. The
latter depicts the trends of the treated unit and the naive average
of the control group. Inspecting the pre-treatment period in the right
panel, the naive average fails to mimic the trend of the treated unit.

\begin{table}
\begin{centering}
\begin{tabular}{cccc}
\toprule 
Years (After Merger) & 2015 & 2016 & 2017\tabularnewline
\midrule 
Treatment Effects & 0.107  & 0.058  & -0.019 \tabularnewline
$p$-Value (block) & 0.037  & 0.074  & 1 \tabularnewline
$p$-Value (i.i.d.) & 0.002  & 0.052  & 0.998 \tabularnewline
\bottomrule
\end{tabular}
\par\end{centering}
\caption{Treatment Effects Averages (by year after the merger)}
\label{tab:avg_effects}
\end{table}

Finally, Table \ref{tab:avg_effects} reports the treatment effects
averaged over each year after the merger. We also report their $p$-values
using the permutation test by \citet{chernozhukov2017exact}. The
advantage of this inferential method in our context is that it does
\emph{not} assume random assignment of the policy intervention unlike
earlier methods that rely on the assumption to ensure the properties
of randomization tests (e.g., \citet{abadie2010synthetic}).\footnote{See \citet{arkhangelsky2019synthetic} for related discussions.}
Consistent with Figure \ref{fig:synth} (left), the short-run effects
in the first and second years are statistically significant. To understand
the magnitude of the effect, note that the standard deviation (SD)
of the gravity measure in the sample is 0.16 as shown in Table \ref{tab:sum_stats}.
Therefore, the treatment effects are on average half of 1 SD. It is
also informative to understand the increase in the measure (roughly
from -9.8 to -9.6) relative to the distribution of the gravity measure
(the left panel of Figure \ref{fig:dist_measures} in the Appendix).
After two years of strong positive effects, the effect of merger dissipates
in the third year. The placebo test results, presented in Table \ref{tab:synth-2}
in the Appendix, show that the treatment effects are not statistically
significant before the merger.

\begin{figure}
\begin{centering}
\includegraphics[scale=0.45]{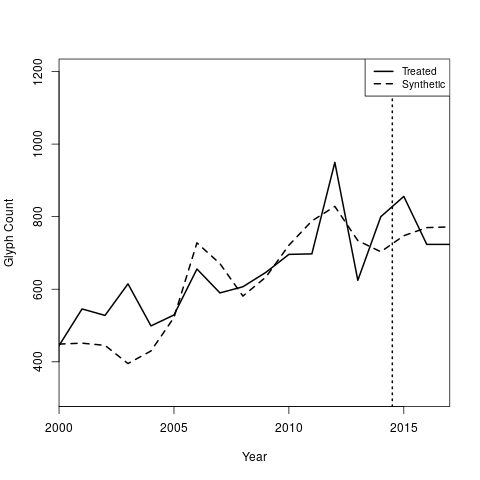}\includegraphics[scale=0.45]{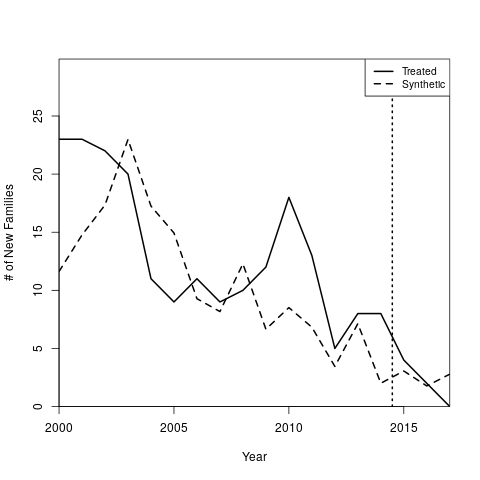}
\par\end{centering}
\caption{Trends of FontFont vs. Synthetic FontFont using Glyph Counts (left)
and the Number of Products (right)}
\medskip{}

{\footnotesize{}}%
\noindent\begin{minipage}[t]{1\columnwidth}%
{\footnotesize{}Note: The solid line depicts the trend of FontFont
and the dashed line depicts the trend of the synthetic control. The
vertical dotted line in each figure indicates the time of the merger.}%
\end{minipage}\label{fig:synth-2}
\end{figure}
We produce analogous results using the mean deviation measure instead
of the gravity measure and find that the results are qualitatively
similar. The effects are positive and statistically significant and
on average larger than 1 SD; see Section \ref{subsec:Mean-Deviation-Measure}
in the Appendix. This suggests that our findings are robust to the
choice of the measure of product differentiation as long as it is
constructed based on the image embeddings. This robustness disappears
if we use more traditional measures for product offerings, such as
the number of products and specifications. Figure \ref{fig:synth-2}
shows that the merger has no effects on both the Glyph counts (which
is the key specification for fonts) and the number of new fonts. Tables
\ref{tab:glyph} and \ref{tab:new_fonts} in the Appendix confirm
that the effects are not statistically significant.

Based on this analysis, we conclude that the merger caused FontFont
to explore a new territory of the product space, at least temporarily.\footnote{Given our data frequency, the foundry's immediate increase in design
differentiation after the merger is feasible because it typically
takes one or two months for a foundry to design a font. } That is, by being part of the parent organization, Monotype, FontFont
increased the visual variety in font design. Before the merger, foundries
owned by Monotype produced fonts and sold them on MyFonts.com. After
the merger, Monotype may have incentives to diversify the product
scope owing to the increased size and efficiency of the firm. It may
also be the case that Monotype tries to avoid cannibalization by spreading
apart its products, thus reducing competition amongst their own foundries.
Such tendency disappears after two years, perhaps because the firm
has either successfully foreclosed the market or found the strategy
unprofitable.\footnote{According to the interview with the employees, we discovered that
the company has gone through a structural change two years after the
merger that is consistent with the second explanation. The details
of the change cannot be publicly revealed.}

\begin{figure}
\begin{centering}
\includegraphics[scale=0.45]{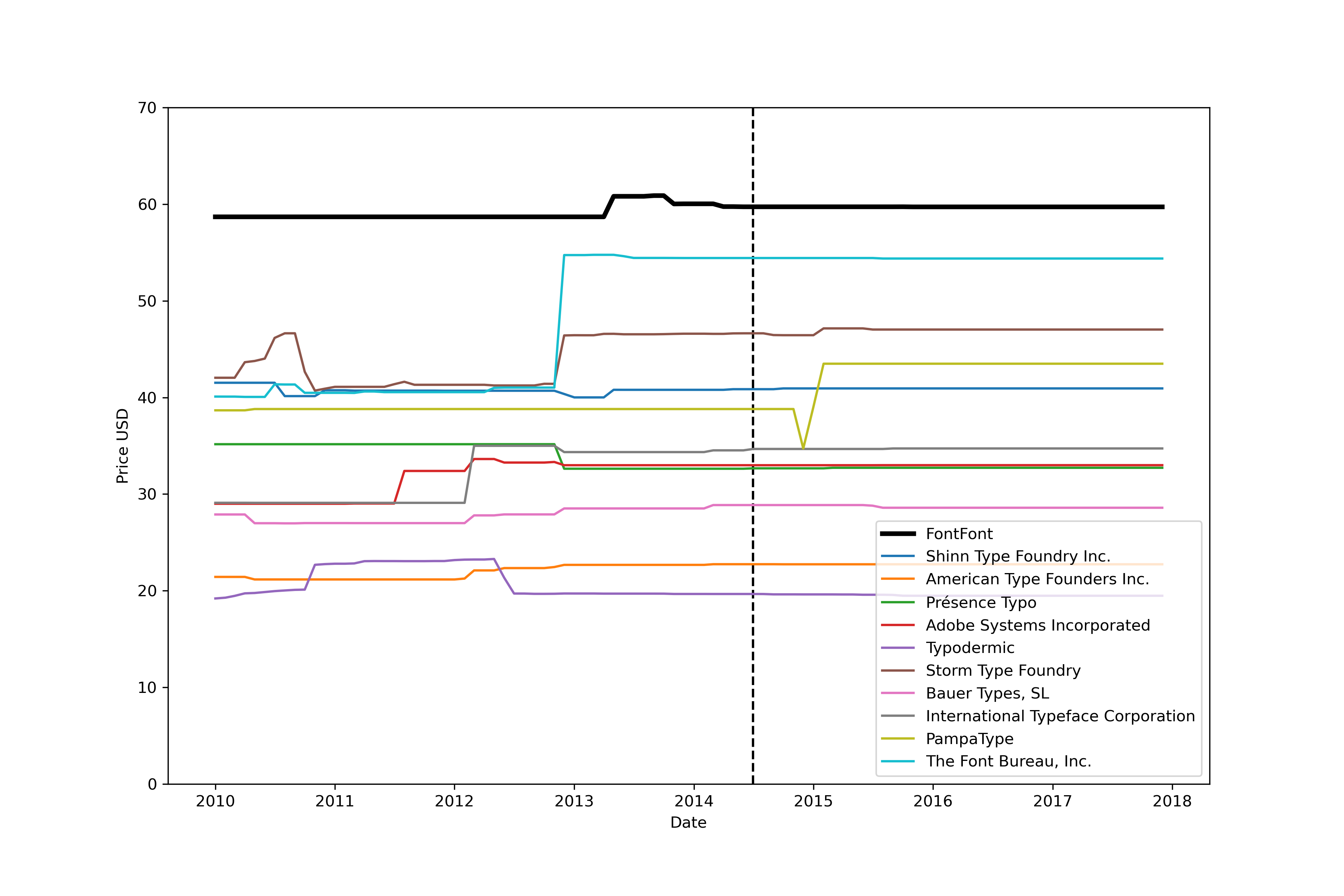}
\par\end{centering}
\caption{Price Trends of FontFont vs. Top Control Foundries}
\medskip{}

{\footnotesize{}}%
\noindent\begin{minipage}[t]{1\columnwidth}%
{\footnotesize{}Note: The price depicted is the average price of all
fonts produced by each foundry in each period. The vertical dotted
line indicates the time of the merger.}%
\end{minipage}\label{fig:price}
\end{figure}

The explanation of our empirical findings echos the prediction under
local monopoly in the simple theoretical model above. Indeed, the
market for fonts may have the aspect of local monopoly in which consumers
exhibit high travel cost. This seems plausible because most of the
consumers in this market are professional designers with sophisticated
tastes for font shapes. The theoretical model also predicts that the
price response to merger would be minimal under local monopoly for
a large range of possible equilibria. Figure \ref{fig:price} corroborates
this theoretical prediction. The figure plots the average prices of
\emph{newly introduced} fonts by FontFont and the top ten control
foundries.\footnote{The range of periods differs from our main analysis due to data limitations.}
The average prices of most of the foundries stay relatively stable,
especially around the time of merger.

\section{Conclusion\label{sec:Conclusion}}

Certain policy questions are better answered by treating high-dimensional,
unstructured attributes as observables and attempting to solve the
resulting dimensionality problem. We propose to quantify the design-oriented
attributes of a product by constructing a low-dimensional product
space of these attributes. We use a modern convolutional neural network
to accomplish this task. We find that these attributes are correlated
with font designers' and consumers' perceptions, as reflected in the
mutual information between tags used to describe the fonts and the
neural network embeddings. We then turn to a causal analysis to understand
the effects of a merger on product differentiation. We find that the
merger increases the product variety in this market. The analytical
framework of this paper can be applied to various products where unstructured
attributes are present.

This paper motivates interesting directions for future research, some
of which may require a structural approach. For structural economic
analyses, we envision a two-step approach of using the embeddings:
construct a neural network embedding to reduce the dimensions of the
product image and then include the embeddings in structural economic
models. Although we could use the neural network to directly predict
economic variables, such as demand and price, the neural network is
not a causal model. Therefore, its counterfactual predictions are
not as credible as those of a more traditional structural model. An
alternative structural approach would be to incorporate dimension
reduction as part of the structural estimation (\citet{chernozhukov2018double},
\citet{foster2019orthogonal}).

Using these structural approaches, we may answer economic questions
related to this market. Related to the economic analyses in this paper,
we may view product differentiation as spatial competition. There
is a close parallel between spatial differentiation and the neural
network embedding product space we constructed. Analogous to \citet{siem2006spatial},
who studies the location choice of video rental stores, we can investigate
the relationship between design choices and the local market power
of the designers. We can also study how third-party and in-house foundries
differ in their product differentiation decisions. One relevant policy
question is the effect of the commission fee of third parties. In
fact, Monotype changed the commission policy during the data collection
period, which may serve as a key policy variation.

Alternatively, we may view product differentiation as intellectual
property. Agents in this industry are subject to license agreements,
which aim to protect the originality of font shapes. Heuristically,
this policy states that ``one cannot produce fonts which shapes are
substantially similar to existing fonts.'' Given the product space
we characterized, one can interpret this policy as imposing a ball
centered around each font, thus preventing other productions within:
producing another font inside the ball is considered a violation.
Then, one can ask what the welfare maximizing level of the policy
(i.e., the optimal radius of the ball) would be and whether the current
level in the market is optimal or suboptimal.

\begin{appendix}

\section{Neural Network Training\label{sec:Neural-Network-Training}}

\subsection{Details of Network Training\label{subsec:Details-of-Network}}

The training iteratively improves the parameters of the network using
batches to estimate the gradient and then update the parameters accordingly
(\citet{wilson2003general}). Each batch contains 270 cropped images
of fonts, or equivalently, 90 triplets. We cropped each image based
on the number of characters in the image.\footnote{For example, the first half of the pangram sentence has 20 characters;
therefore, to crop 5 characters, we would take 40 percent of the pixels
in the first half of the image. We tried crops with 3, 4, 6, and 7
different characters. We also tried different cropping schemes such
as using the white space between characters.} As the gradient is evaluated at more batches, the parameters in the
network are adjusted. The number of trainable parameters are $90,000=3\times(3\times100\times100)$
(layers $\times$ input size). The training of the network is completed
when the loss function reaches below 0.7. The learning rate is initially
0.05, and then lower to 0.01 to finalized the model. Here are the
remaining hyper-parameter values: the batch size is 90, epoch size
500, weight decay $1\times10^{-4}$, and the margin $\alpha$ is 0.2.
The training time took approximately 24 hours with 4 GPUs (Nvidia
1080-TI).

\subsection{Evaluation of Trained Network\label{sec:Internal-Evaluation}}

To evaluate the neural network embeddings, we create test sets of
triplets that the neural network has never seen during the training.
Using the test sets, the task is to identify whether the image in
a triplet is a positive or a negative. It is classified as a positive
if its Euclidean distance from the anchor in the trained embedding
space is less than a pre-defined threshold (via cross validation)
and a negative otherwise. We describe the results of tests on two
different test sets. The first test set, ``easy,'' randomly samples
pairs of a positive crop within the same family as an anchor and a
negative crop from different families, for which the performance is
evaluated. The second test set, ``hard,'' samples pairs of a positive
crop within the same style of a family as an anchor and a negative
crop from different styles within the same family.

The analysis involves true positives (TP), true negatives (TN), false
positives (FP), and false negatives (FN). Table \ref{tab:scalar}
presents the accuracy and validation rates for the neural network
in each of these test sets with different crop sizes. The accuracy
is a measure of how well the neural network embeddings perform in
the classification task. Overall, it gets about 90 percent of the
triplets correct.

\begin{figure}
\noindent \begin{centering}
\begin{tabular}{cc}
\toprule 
``Easy'' Test Set & ``Hard'' Test Set\tabularnewline
\midrule 
\includegraphics[scale=0.5]{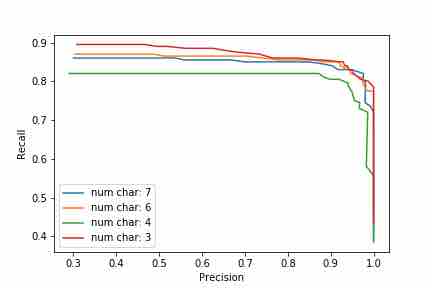} & \includegraphics[scale=0.5]{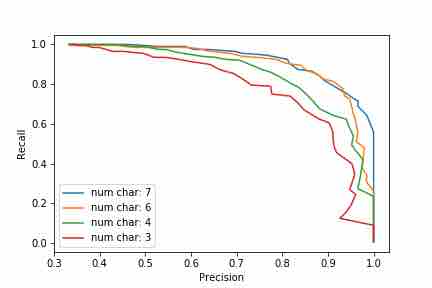}\tabularnewline
\bottomrule
\end{tabular}
\par\end{centering}
\caption{Precision Recall Curves}

\label{fig:PR_curve}
\end{figure}

We also analyze the trade-off between Type-I and Type-II errors in
classification. The validation rate is the true acceptance rate and
is calculated as TP/(TP+FN). The false acceptance rate (FAR) is given
by FP/(TN+FP). In addition, we plot precision and recall curves. Precision
is related to the Type-I errors and is calculated as TP/(TP+FP). Recall
is related to the Type-II errors and is given by the formula TP/(TP+FN).
The precision recall curve in Figure \ref{fig:PR_curve} shows the
trade-off between these types of errors. There exists a steeper trade-off
between precision and recall in the hard test set.
\begin{table}
\centering{}%
\begin{tabular}{ccccc}
\toprule 
Crop Size (Characters) & Test Set & Accuracy & Validation Rate & FAR\tabularnewline
\midrule 
7 & Hard & 0.8925 & 0.09375 & 0\tabularnewline
7 & Easy & 0.8975 & 0.47875 & 0\tabularnewline
6 & Hard & 0.8825 & 0.04875 & 0\tabularnewline
6 & Easy & 0.89667 & 0.53875 & 0\tabularnewline
4 & Hard & 0.86333 & 0.02 & 0\tabularnewline
4 & Easy & 0.8925 & 0.46875 & 0\tabularnewline
3 & Hard & 0.76417 & 0.00875 & 0\tabularnewline
3 & Easy & 0.88167 & 0.48625 & 0\tabularnewline
\bottomrule
\end{tabular}\caption{Internal Validation by Accuracy, Validation Rate, and FAR}
\label{tab:scalar}
\end{table}

Overall, all the statistics we report here show that our neural network
performs well. The low validation rate in the hard test set suggests
that there is room for improvement in differentiating between different
styles.

\section{Trend Analyses\label{sec:Trend-Analysis}}

To further illustrate the usefulness of the embeddings, we analyze
the supply- and demand-side trends in font style. This also provides
an additional background for the causal analysis of the merger in
Section \ref{sec:Effects-of-Merger}. On the supply side there are
constant entries of new products in the marketplace. On the demand
side, on average, more than 1,000 fonts are (stably) sold per day.
Of course, demand and supply are endogenous, so this analysis is only
a descriptive analysis. We, however, believe it reveals some interesting
patterns in this market. 

\subsection{Supply-Side Trend}

\begin{figure}
\noindent \begin{centering}
\includegraphics[scale=0.34]{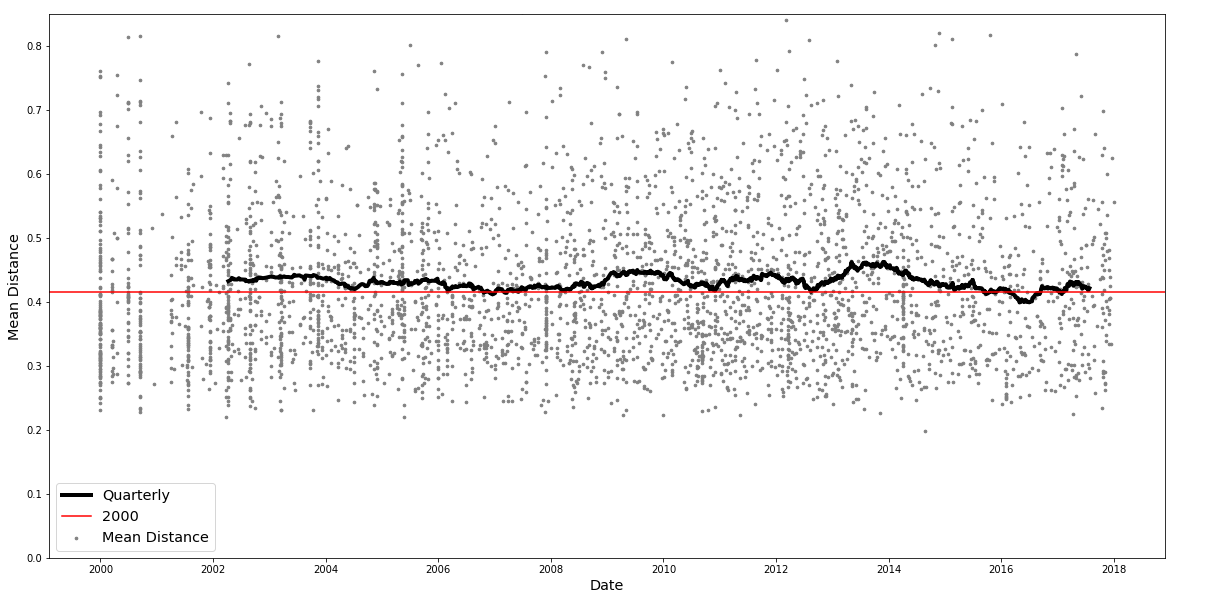}
\par\end{centering}
\caption{Trends in Mean Distance of Incumbents ($\le2000$, in red) and New
Entrants ($2001$--$2017$)}
\label{supply_trend}
\end{figure}
We analyze the supply-side trend in the Euclidean distance from Averia.
We are particularly interested in how the style of fonts newly entering
the market differs from the style of incumbent fonts. In Figure \ref{supply_trend},
each dot represents the daily mean distance from Averia for a range
of periods. The red horizontal line is the mean distance of \textit{all}
fonts introduced before 2001, which we view as incumbents. The black
line depicts the quarterly moving averages of the daily mean for fonts
newly introduced since 2001 each day, which we view as entrants. Overall,
we find that entrants have font shapes that are more experimental
or innovative than those of incumbents, possibly to avoid competition
and establish market power distant from the incumbents in the product
space.

\subsection{Demand-Side Trend}

\begin{figure}
\begin{centering}
\includegraphics[scale=0.25]{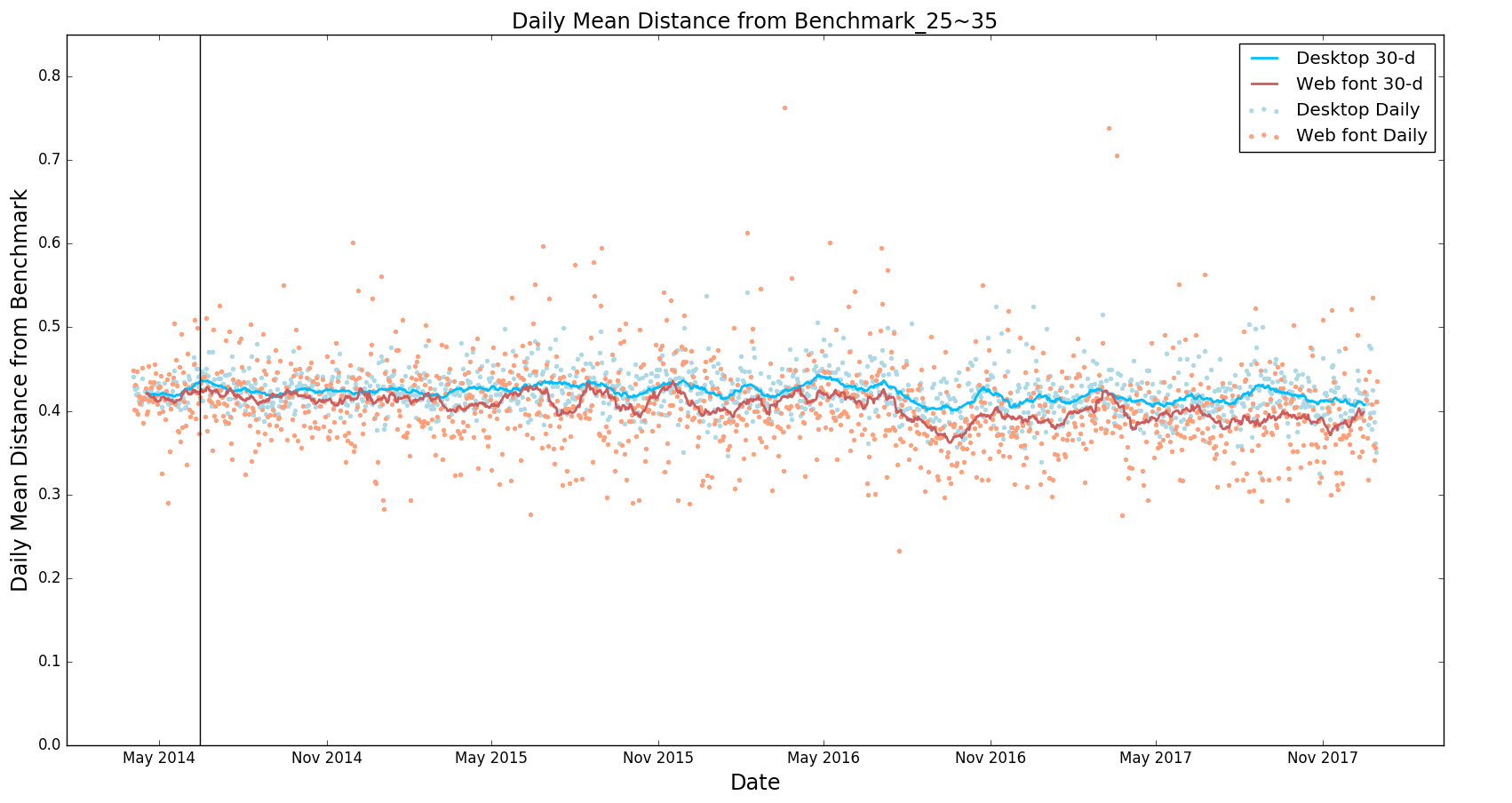}
\par\end{centering}
\begin{centering}
\includegraphics[scale=0.25]{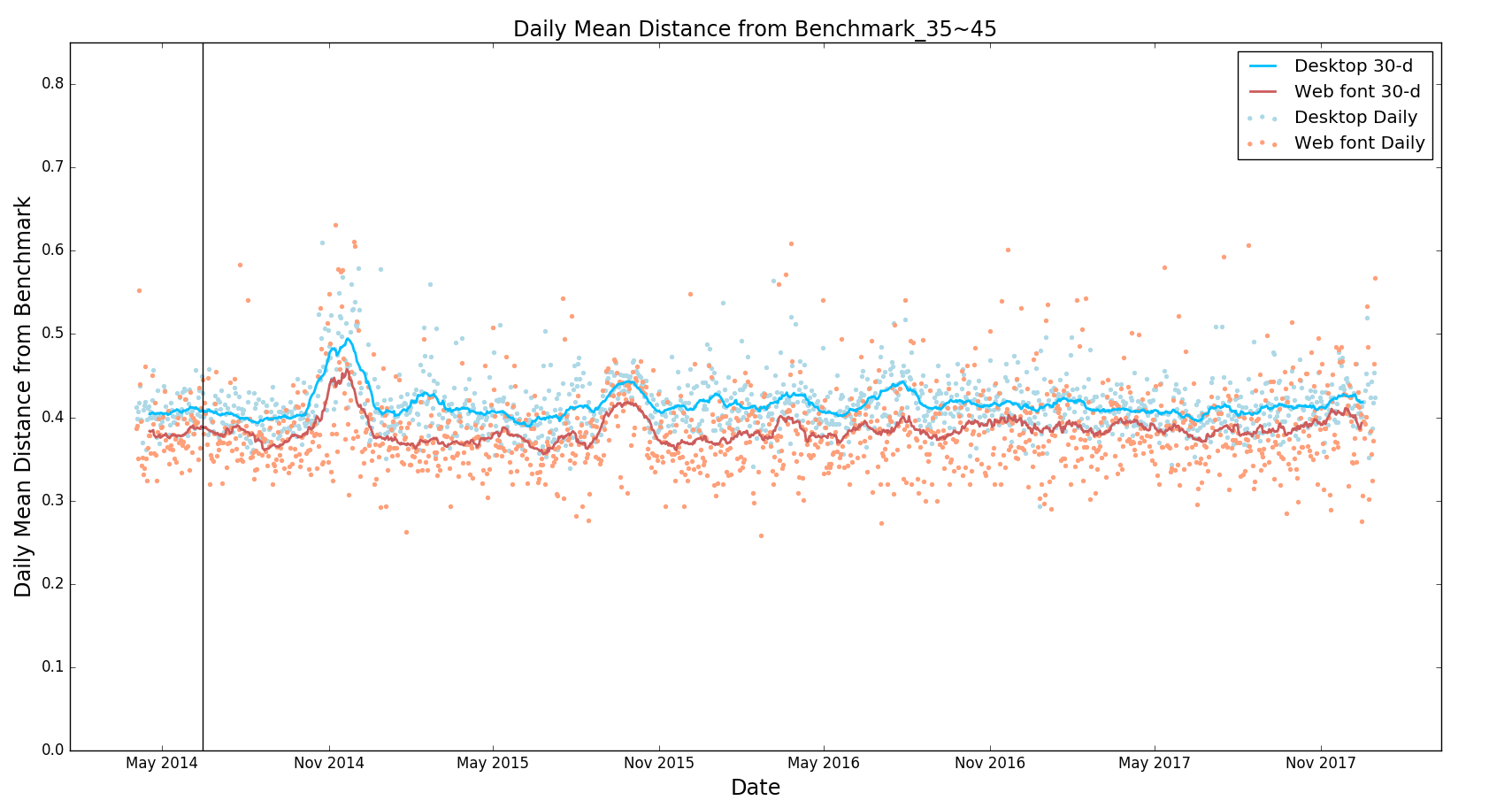}
\par\end{centering}
\caption{Trends in (Weighted) Mean Distance of Purchased Fonts (for USD 25--35
and USD 35--45, by license types)}
\medskip{}

{\footnotesize{}}%
\noindent\begin{minipage}[t]{1\columnwidth}%
{\footnotesize{}Note: The vertical line in each figure indicates the
time of the merger we analyze in Section \ref{sec:Effects-of-Merger}.}%
\end{minipage}\label{conditional_trend}
\end{figure}
We now analyze the trend in the Euclidean distance between Averia
and fonts purchased by consumers between 2014 and 2017.\footnote{The time span is shorter than that for the supply-side analysis due
to the missing license type data for earlier periods.} To remove the supply-driven factor, we condition on price and focus
on the price ranges of USD 25--35 and USD 35--45, where promotions
are rare. These price ranges are the most common in the market. Figure
\ref{conditional_trend} plots the trend conditional on the price
being USD 25--35 and USD 35--45. We also plot the trend for the
two major license types: desktop license and web font license. Again,
each dot represents the daily mean distance weighted by the number
of purchases. We also plot monthly moving averages.

Interestingly, in both figures, we find that fonts sold under desktop
license are more experimental (or less conservative) than fonts sold
under web font license. Because both licenses are offered for most
fonts, this difference cannot be attributed to the supply-side decision
but is the result of consumer decisions.

\subsection{Summary of Findings}

Based on our analyses, we find the following stylized facts. (i) Entrants
are more innovative than incumbents. (ii) Consumer preferences are
stable during 2014, the year that witnessed the merger of our interest.\footnote{This feature is uniformly found in all price ranges, but we do not
report the results for succinctness.} Based on this finding, we assume that the demand side has a negligible
influence on the change in product differentiation decisions by the
merging firm around the time of merger. (iii) Consumers have different
preferences over shapes depending on the license type they purchase.
Consumers prefer more conservative shapes for web font licenses and
more experimental shapes for desktop licenses. Usually, web font licenses
are used on webpages, where legibility is important, whereas desktop
licenses are used in printed material (e.g., posters, cards), where
designers (as consumers) have more control over the design environment.

\section{Supplemental Findings for Merger Analysis\label{sec:Supplemental-Findings-for}}

\subsection{Placebo Test with Gravity Measure\label{subsec:Gravity-Measure}}

\begin{table}[H]
\begin{centering}
\begin{tabular}{ccccccc}
\toprule 
{} & 2003 & 2004 & 2005 & 2006 & 2007 & 2008\tabularnewline
\midrule 
Treatment Effects & -0.0172 & -0.0158 & -0.0298 & 0.013 & 0.0473 & 0.0351\tabularnewline
$p$-Value (block)  & 0.75 & 0.1667 & 1 & 0.5 & 0.0833 & 0.9286\tabularnewline
$p$-Value (i.i.d.)  & 0.6577 & 0.0654 & 0.9672 & 0.5029 & 0.0668 & 0.9376\tabularnewline
\bottomrule
\end{tabular}
\par\end{centering}
\begin{centering}
\begin{tabular}{ccccccc}
\toprule 
{} & 2009 & 2010 & 2011 & 2012 & 2013 & 2014\tabularnewline
\midrule 
Treatment Effects & -0.0111 & -0.0288 & 0.012 & 0.0727 & -0.0316 & -0.0659\tabularnewline
$p$-Value (block)  & 0.8125 & 0.6667 & 0.2 & 0.3636 & 0.1667 & 0.5385\tabularnewline
$p$-Value (i.i.d.)  & 0.873 & 0.5351 & 0.1068 & 0.2523 & 0.1452 & 0.5335\tabularnewline
\bottomrule
\end{tabular}
\par\end{centering}
\caption{Placebo Test: Treatment Effects Before Merger (Using Gravity Measure)}

\centering{}\label{tab:synth-2}
\end{table}

\begin{figure}
\begin{centering}
\includegraphics[scale=0.4]{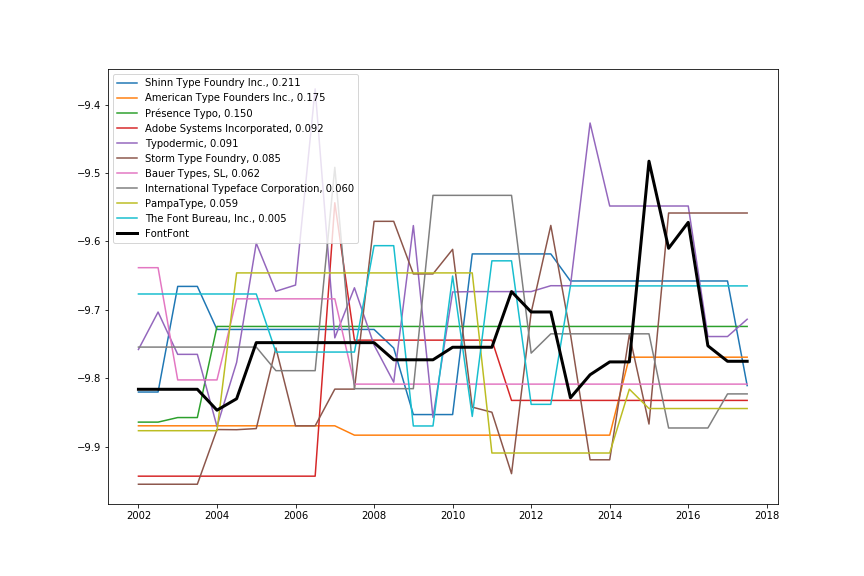}
\par\end{centering}
\caption{Trends of FontFont vs. Top 10 Control Units (Using Gravity Measure)}
\medskip{}

{\footnotesize{}}%
\noindent\begin{minipage}[t]{1\columnwidth}%
{\footnotesize{}Note: The values next to the foundry names indicate
the weights used to construct the synthetic control.}%
\end{minipage}\label{fig:synth-gravity}
\end{figure}

\subsection{Merger Effects and Placebo Test with Mean Deviation Measure\label{subsec:Mean-Deviation-Measure}}

\begin{figure}
\begin{centering}
\includegraphics[scale=0.45]{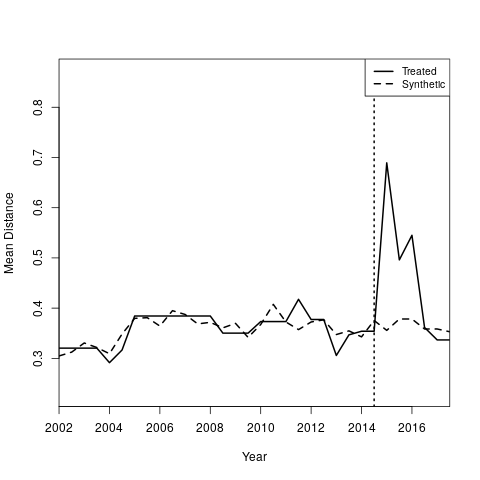}\includegraphics[scale=0.45]{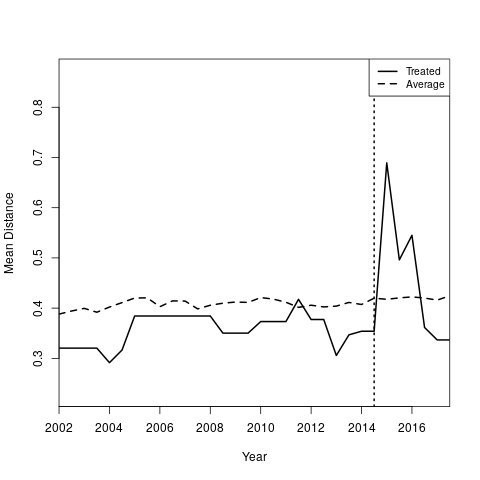}
\par\end{centering}
\caption{Trends of FontFont vs. Synthetic FontFont (left) and Naive Control
Group (right)---Using Mean Deviation Measure}
\medskip{}

{\footnotesize{}}%
\noindent\begin{minipage}[t]{1\columnwidth}%
{\footnotesize{}Note: The solid line depicts the trend of FontFont
and the dashed line depicts the trend of the synthetic control (left)
or the naive average trend among all the control units (right). The
vertical dotted line in each figure indicates the time of the merger.}%
\end{minipage}\label{fig:synth-1}
\end{figure}

\begin{table}[H]
\begin{centering}
\begin{tabular}{ccccccc}
\toprule 
{} & 2003 & 2004 & 2005 & 2006 & 2007 & 2008\tabularnewline
\midrule 
Treatment Effects & -0.0036 & -0.0095 & 0.0177 & 0.0241 & 0.0388 & 0.005\tabularnewline
$p$-Value (block)  & 0.75 & 0.3333 & 0.25 & 0.3 & 0.1667 & 0.7857\tabularnewline
$p$-Value (i.i.d.)  & 0.8292 & 0.3391 & 0.1486 & 0.1494 & 0.0836 & 0.7552\tabularnewline
\bottomrule
\end{tabular}
\par\end{centering}
\begin{centering}
\begin{tabular}{ccccccc}
\toprule 
{} & 2009 & 2010 & 2011 & 2012 & 2013 & 2014\tabularnewline
\midrule 
Treatment Effects & -0.0444 & -0.0224 & 0.0371 & 0.0323 & -0.0144 & -0.0056\tabularnewline
$p$-Value (block)  & 0.125 & 0.7222 & 0.1 & 0.8182 & 0.625 & 0.7308\tabularnewline
$p$-Value (i.i.d.)  & 0.096 & 0.6569 & 0.0764 & 0.803 & 0.5563 & 0.7047\tabularnewline
\bottomrule
\end{tabular}
\par\end{centering}
\caption{Placebo Test: Treatment Effects Before Merger (Using Mean Deviation)}

\centering{}\label{tab:synth-2-1}
\end{table}

\begin{table}[H]
\begin{centering}
\begin{tabular}{cccc}
\toprule 
{} & 2015 & 2016 & 2017\tabularnewline
\midrule 
Treatment Effects & 0.1412 & 0.061 & -0.0305\tabularnewline
$p$-Value (block)  & 0.037 & 0.0741 & 0.1111\tabularnewline
$p$-Value (i.i.d.)  & 0.0034 & 0.0704 & 0.0716\tabularnewline
\bottomrule
\end{tabular}
\par\end{centering}
\caption{Treatment Effects After Merger (Using Mean Deviation)}

\centering{}\label{tab:synth-1-1}
\end{table}

\begin{figure}
\begin{centering}
\includegraphics[scale=0.4]{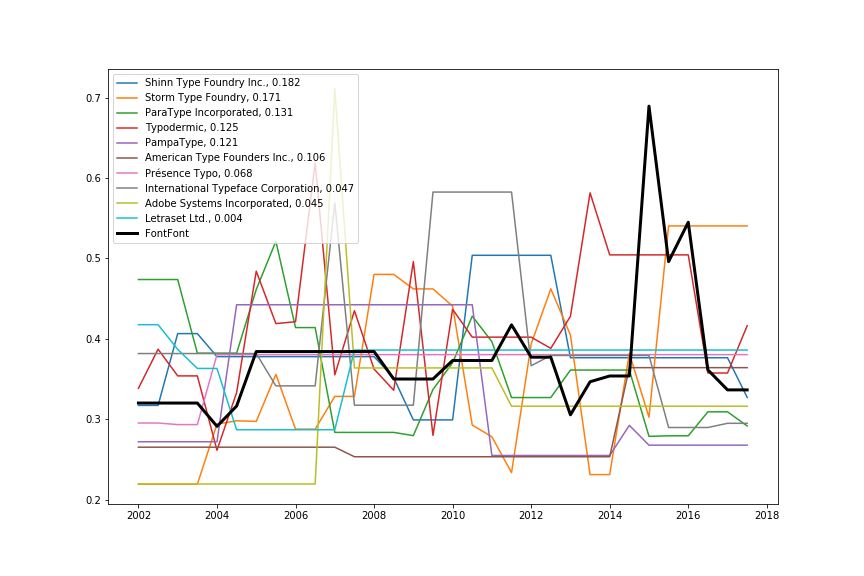}
\par\end{centering}
\caption{Trends of FontFont vs. Top 10 Control Units (Using Mean Deviation)}
\medskip{}

{\footnotesize{}}%
\noindent\begin{minipage}[t]{1\columnwidth}%
{\footnotesize{}Note: The values next to the foundry names indicate
the weights used to construct the synthetic control.}%
\end{minipage}\label{fig:synth-mean_dev}
\end{figure}

\subsection{Merger Effects with Traditional Measures of Product Offerings\label{subsec:Traditional-Measures-of}}

\subsubsection{Glyph Counts}

\begin{table}[H]
\begin{centering}
\begin{tabular}{ccccccc}
\toprule 
{} & 2003 & 2004 & 2005 & 2006 & 2007 & 2008\tabularnewline
\midrule 
Treatment Effects  & 125.1264 & 48.1483 & -19.9343 & -14.6911 & -57.5153 & 1.1326\tabularnewline
$p$-Value (block)  & 0.25 & 1 & 0.8333 & 0.8571 & 0.375 & 0.4444\tabularnewline
$p$-Value (i.i.d.)  & 0.2376 & 1 & 0.829 & 0.8534 & 0.3759 & 0.4469\tabularnewline
\bottomrule
\end{tabular}
\par\end{centering}
\begin{centering}
\begin{tabular}{ccccccc}
\toprule 
{} & 2009 & 2010 & 2011 & 2012 & 2013 & 2014\tabularnewline
\midrule 
Treatment Effects  & 29.1529 & -23.4679 & 29.4617  & 178.8836  & 99.5249  & 11.8745 \tabularnewline
$p$-Value (block)  & 0.7 & 0.5455 & 0.25  & 0.0769  & 0.5714  & 0.4 \tabularnewline
$p$-Value (i.i.d.)  & 0.6991 & 0.5409 & 0.2579  & 0.0786  & 0.5749  & 0.4063 \tabularnewline
\bottomrule
\end{tabular}
\par\end{centering}
\caption{Placebo Test: Treatment Effects Before Merger (Using Glyph Counts)}
\end{table}

\begin{table}[H]
\begin{centering}
\begin{tabular}{cccc}
\toprule 
{} & 2015 & 2016 & 2017\tabularnewline
\midrule 
Treatment Effects  & 127.9152 & 74.8633 & 73.8973\tabularnewline
$p$-Value (block)  & 0.25 & 0.75 & 0.75\tabularnewline
$p$-Value (i.i.d.)  & 0.2645 & 0.7395 & 0.7574\tabularnewline
\bottomrule
\end{tabular}
\par\end{centering}
\caption{Treatment Effects After Merger (Using Glyph Counts)}
\label{tab:glyph}
\end{table}

\subsubsection{Number of New Fonts}

\begin{table}[H]
\begin{centering}
\begin{tabular}{ccccccc}
\toprule 
{} & 2003 & 2004 & 2005 & 2006 & 2007 & 2008\tabularnewline
\midrule 
Treatment Effects  & 48.5002 & 1.0007 & 16.0004 & 232 & 248 & 156\tabularnewline
$p$-Value (block)  & 1 & 0.8 & 0.6667 & 0.1429 & 0.625 & 0.2222\tabularnewline
$p$-Value (i.i.d.)  & 1 & 0.8034 & 0.6677 & 0.1408 & 0.6241 & 0.229\tabularnewline
\bottomrule
\end{tabular}
\par\end{centering}
\begin{centering}
\begin{tabular}{ccccccc}
\toprule 
{} & 2009 & 2010 & 2011 & 2012 & 2013 & 2014\tabularnewline
\midrule 
Treatment Effects  & 64.5 & 210.0034 & 376.5  & 144.5  & 164  & 155 \tabularnewline
$p$-Value (block)  & 0.4 & 0.0909 & 0.4167  & 0.6154  & 0.2857  & 0.7333 \tabularnewline
$p$-Value (i.i.d.)  & 0.3939 & 0.0938 & 0.4115  & 0.6137  & 0.2901  & 0.7297 \tabularnewline
\bottomrule
\end{tabular}
\par\end{centering}
\caption{Placebo Test: Treatment Effects Before Merger (Using Number of New
Fonts)}
\end{table}

\begin{table}[H]
\begin{centering}
\begin{tabular}{cccc}
\toprule 
{} & 2015 & 2016 & 2017\tabularnewline
\midrule 
Treatment Effects  & 73.5002 & 29.0002 & 32.0001\tabularnewline
$p$-Value (block)  & 0.5625 & 0.9375 & 1\tabularnewline
$p$-Value (i.i.d.)  & 0.5599 & 0.9426 & 1\tabularnewline
\bottomrule
\end{tabular}
\par\end{centering}
\caption{Treatment Effects After Merger (Using Number of New Fonts)}
\label{tab:new_fonts}
\end{table}

\subsection{Distributions of the Measures\label{subsec:dist_measures}}

\begin{figure}
\begin{centering}
\includegraphics[scale=0.55]{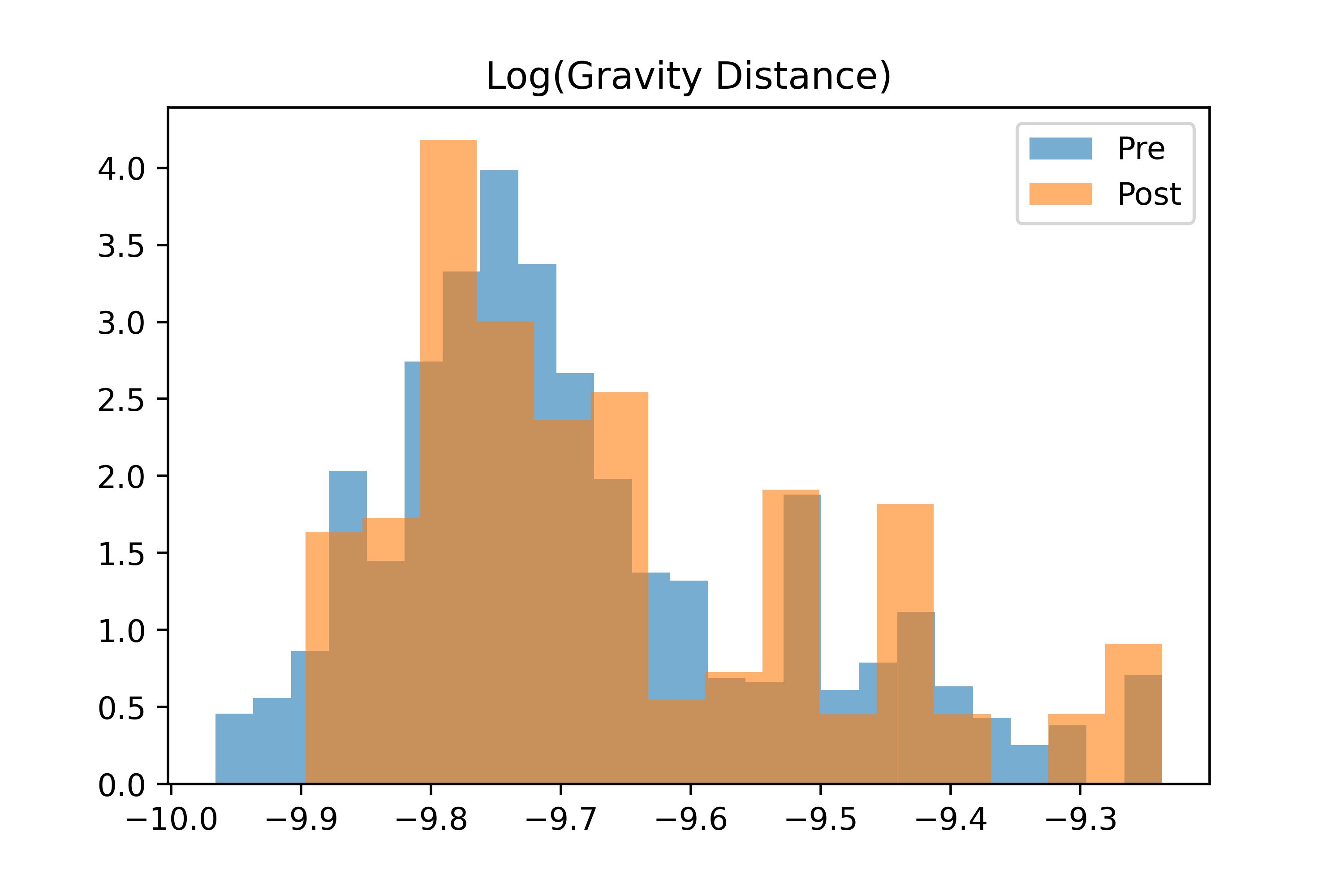}\includegraphics[scale=0.55]{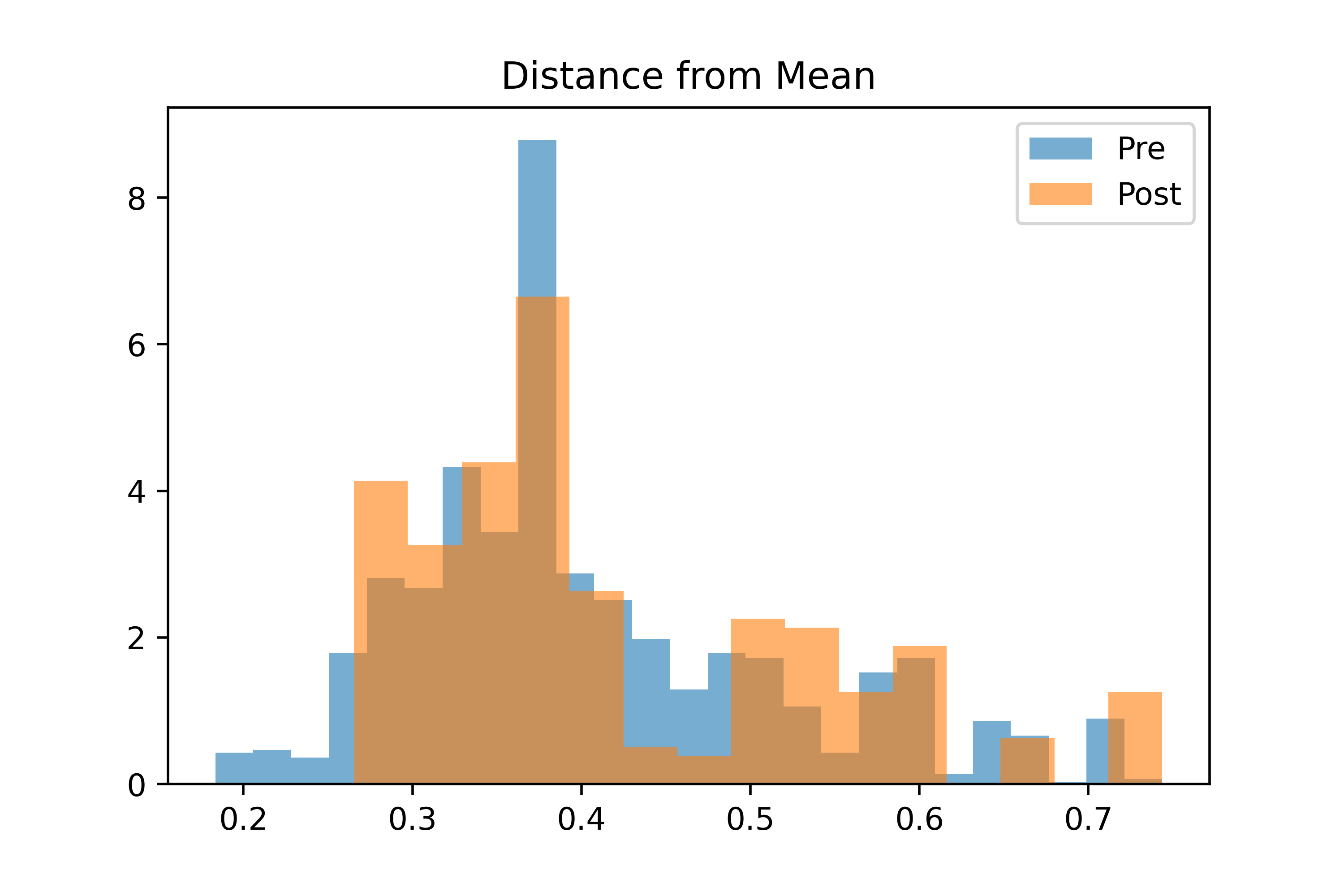}
\par\end{centering}
\caption{Histograms of the Gravity Measure (left) and Mean Deviation Measure
(right), Before and After the Merger}

\label{fig:dist_measures}
\end{figure}

Figure \ref{fig:dist_measures} presents the distributions of the
gravity and mean deviation measures before and after the merger. It
is helpful to understand the magnitude of the estimates relative to
the distributions.

\end{appendix}

\bibliographystyle{ecta}
\bibliography{FontMkt}

\begin{thebibliography}{64}
\newcommand{\enquote}[1]{``#1''}
\expandafter\ifx\csname natexlab\endcsname\relax\def\natexlab#1{#1}\fi

\bibitem[\protect\citeauthoryear{Abadie}{Abadie}{2021}]{abadie2021using}
\textsc{Abadie, A.} (2021): \enquote{Using synthetic controls: Feasibility,
  data requirements, and methodological aspects,} \emph{Journal of Economic
  Literature}, 59, 391--425.

\bibitem[\protect\citeauthoryear{Abadie, Diamond, and Hainmueller}{Abadie
  et~al.}{2010}]{abadie2010synthetic}
\textsc{Abadie, A., A.~Diamond, and J.~Hainmueller} (2010): \enquote{Synthetic
  control methods for comparative case studies: Estimating the effect of
  California's tobacco control program,} \emph{Journal of the American
  statistical Association}, 105, 493--505.

\bibitem[\protect\citeauthoryear{Abadie and Gardeazabal}{Abadie and
  Gardeazabal}{2003}]{abadie2003economic}
\textsc{Abadie, A. and J.~Gardeazabal} (2003): \enquote{The economic costs of
  conflict: A case study of the Basque Country,} \emph{American economic
  review}, 93, 113--132.

\bibitem[\protect\citeauthoryear{Al-Halah, Stiefelhagen, and Grauman}{Al-Halah
  et~al.}{2017}]{al2017fashion}
\textsc{Al-Halah, Z., R.~Stiefelhagen, and K.~Grauman} (2017): \enquote{Fashion
  forward: Forecasting visual style in fashion,} in \emph{Proceedings of the
  IEEE International Conference on Computer Vision}, 388--397.

\bibitem[\protect\citeauthoryear{Angrist and Pischke}{Angrist and
  Pischke}{2010}]{angrist2010credibility}
\textsc{Angrist, J.~D. and J.-S. Pischke} (2010): \enquote{The credibility
  revolution in empirical economics: How better research design is taking the
  con out of econometrics,} \emph{Journal of economic perspectives}, 24, 3--30.

\bibitem[\protect\citeauthoryear{Arkhangelsky, Athey, Hirshberg, Imbens, and
  Wager}{Arkhangelsky et~al.}{2019}]{arkhangelsky2019synthetic}
\textsc{Arkhangelsky, D., S.~Athey, D.~A. Hirshberg, G.~W. Imbens, and
  S.~Wager} (2019): \enquote{Synthetic difference in differences,}
  \emph{National Bureau of Economic Research}.

\bibitem[\protect\citeauthoryear{Ashenfelter and Hosken}{Ashenfelter and
  Hosken}{2008}]{ashenfelter2008effect}
\textsc{Ashenfelter, O. and D.~Hosken} (2008): \enquote{The effect of mergers
  on consumer prices: Evidence from five selected case studies,} \emph{National
  Bureau of Economic Research}.

\bibitem[\protect\citeauthoryear{Atalay, Sorensen, Zhu, and Sullivan}{Atalay
  et~al.}{2020}]{atalay2020post}
\textsc{Atalay, E., A.~Sorensen, W.~Zhu, and C.~Sullivan} (2020):
  \enquote{Post-Merger Product Repositioning: An Empirical Analysis,} \emph{FRB
  of Philadelphia Working Paper}.

\bibitem[\protect\citeauthoryear{Bajari and Benkard}{Bajari and
  Benkard}{2005}]{bajari2005demand}
\textsc{Bajari, P. and C.~L. Benkard} (2005): \enquote{Demand estimation with
  heterogeneous consumers and unobserved product characteristics: A hedonic
  approach,} \emph{Journal of political economy}, 113, 1239--1276.

\bibitem[\protect\citeauthoryear{Bajari, Cen, Chernozhukov, Manukonda, Wang,
  Huerta, Li, Leng, Monokroussos, Vijaykunar et~al.}{Bajari
  et~al.}{2021}]{bajari2018hedonic}
\textsc{Bajari, P., Z.~Cen, V.~Chernozhukov, M.~Manukonda, J.~Wang, R.~Huerta,
  J.~Li, L.~Leng, G.~Monokroussos, S.~Vijaykunar, et~al.} (2021):
  \enquote{Hedonic prices and quality adjusted price indices powered by AI,}
  \emph{cemmap working paper}.

\bibitem[\protect\citeauthoryear{Berry, Levinsohn, and Pakes}{Berry
  et~al.}{1995}]{berry1995automobile}
\textsc{Berry, S., J.~Levinsohn, and A.~Pakes} (1995): \enquote{Automobile
  prices in market equilibrium,} \emph{Econometrica: Journal of the Econometric
  Society}, 841--890.

\bibitem[\protect\citeauthoryear{Berry and Waldfogel}{Berry and
  Waldfogel}{2001}]{berry2001}
\textsc{Berry, S.~T. and J.~Waldfogel} (2001): \enquote{{Do Mergers Increase
  Product Variety? Evidence from Radio Broadcasting},} \emph{The Quarterly
  Journal of Economics}, 116, 1009--1025.

\bibitem[\protect\citeauthoryear{Bottou}{Bottou}{2010}]{bottou2010large}
\textsc{Bottou, L.} (2010): \enquote{Large-scale machine learning with
  stochastic gradient descent,} in \emph{Proceedings of COMPSTAT 2010},
  Springer, 177--186.

\bibitem[\protect\citeauthoryear{Burnap, Liu, Pan, Lee, Gonzalez, and
  Papalambros}{Burnap et~al.}{2016}]{burnap2016estimating}
\textsc{Burnap, A., Y.~Liu, Y.~Pan, H.~Lee, R.~Gonzalez, and P.~Y. Papalambros}
  (2016): \enquote{Estimating and exploring the product form design space using
  deep generative models,} in \emph{International Design Engineering Technical
  Conferences and Computers and Information in Engineering Conference},
  American Society of Mechanical Engineers, vol. 50107, V02AT03A013.

\bibitem[\protect\citeauthoryear{Campbell and Kautz}{Campbell and
  Kautz}{2014}]{campbell2014learning}
\textsc{Campbell, N.~D. and J.~Kautz} (2014): \enquote{Learning a manifold of
  fonts,} \emph{ACM Transactions on Graphics (TOG)}, 33, 91.

\bibitem[\protect\citeauthoryear{Chernozhukov, Chetverikov, Demirer, Duflo,
  Hansen, Newey, and Robins}{Chernozhukov
  et~al.}{2018}]{chernozhukov2018double}
\textsc{Chernozhukov, V., D.~Chetverikov, M.~Demirer, E.~Duflo, C.~Hansen,
  W.~Newey, and J.~Robins} (2018): \enquote{Double/debiased machine learning
  for treatment and structural parameters,} \emph{Economics Journal}, 21,
  C1--C68.

\bibitem[\protect\citeauthoryear{Chernozhukov, W{\"u}thrich, and
  Zhu}{Chernozhukov et~al.}{2021}]{chernozhukov2017exact}
\textsc{Chernozhukov, V., K.~W{\"u}thrich, and Y.~Zhu} (2021): \enquote{An
  exact and robust conformal inference method for counterfactual and synthetic
  controls,} \emph{Journal of the American Statistical Association}, 116,
  1849--1864.

\bibitem[\protect\citeauthoryear{Dosovitskiy, Springenberg, Tatarchenko, and
  Brox}{Dosovitskiy et~al.}{2016}]{dosovitskiy2016learning}
\textsc{Dosovitskiy, A., J.~T. Springenberg, M.~Tatarchenko, and T.~Brox}
  (2016): \enquote{Learning to generate chairs, tables and cars with
  convolutional networks,} \emph{IEEE transactions on pattern analysis and
  machine intelligence}, 39, 692--705.

\bibitem[\protect\citeauthoryear{Economides}{Economides}{1989}]{economides1989}
\textsc{Economides, N.} (1989): \enquote{Symmetric equilibrium existence and
  optimality in differentiated product markets,} \emph{Journal of Economic
  Theory}, 47, 178--194.

\bibitem[\protect\citeauthoryear{Fan}{Fan}{2013}]{fan2013}
\textsc{Fan, Y.} (2013): \enquote{Ownership Consolidation and Product
  Characteristics: A Study of the US Daily Newspaper Market,} \emph{American
  Economic Review}, 103, 1598--1628.

\bibitem[\protect\citeauthoryear{Fan and Yang}{Fan and
  Yang}{2020}]{fan2020competition}
\textsc{Fan, Y. and C.~Yang} (2020): \enquote{Competition, product
  proliferation, and welfare: A study of the US smartphone market,}
  \emph{American Economic Journal: Microeconomics}, 12, 99--134.

\bibitem[\protect\citeauthoryear{Foster and Syrgkanis}{Foster and
  Syrgkanis}{2019}]{foster2019orthogonal}
\textsc{Foster, D.~J. and V.~Syrgkanis} (2019): \enquote{Orthogonal statistical
  learning,} \emph{arXiv preprint arXiv:1901.09036}.

\bibitem[\protect\citeauthoryear{Friedman, Hastie, and Tibshirani}{Friedman
  et~al.}{2001}]{friedman2001elements}
\textsc{Friedman, J., T.~Hastie, and R.~Tibshirani} (2001): \emph{The elements
  of statistical learning}, vol.~1, Springer series in statistics New York.

\bibitem[\protect\citeauthoryear{Galenson and Weinberg}{Galenson and
  Weinberg}{2000}]{galenson2000age}
\textsc{Galenson, D.~W. and B.~A. Weinberg} (2000): \enquote{Age and the
  quality of work: The case of modern American painters,} \emph{Journal of
  Political Economy}, 108, 761--777.

\bibitem[\protect\citeauthoryear{Galenson and Weinberg}{Galenson and
  Weinberg}{2001}]{galenson2001creating}
---\hspace{-.1pt}---\hspace{-.1pt}--- (2001): \enquote{Creating modern art: The
  changing careers of painters in France from impressionism to cubism,}
  \emph{American Economic Review}, 91, 1063--1071.

\bibitem[\protect\citeauthoryear{Gentzkow, Kelly, and Taddy}{Gentzkow
  et~al.}{2019{\natexlab{a}}}]{gentzkow2019}
\textsc{Gentzkow, M., B.~Kelly, and M.~Taddy} (2019{\natexlab{a}}):
  \enquote{Text as Data,} \emph{Journal of Economic Literature}, 57, 535--74.

\bibitem[\protect\citeauthoryear{Gentzkow, Shapiro, and Taddy}{Gentzkow
  et~al.}{2019{\natexlab{b}}}]{gentzkow2019measuring}
\textsc{Gentzkow, M., J.~M. Shapiro, and M.~Taddy} (2019{\natexlab{b}}):
  \enquote{Measuring group differences in high-dimensional choices: method and
  application to congressional speech,} \emph{Econometrica}, 87, 1307--1340.

\bibitem[\protect\citeauthoryear{Glaeser, Kominers, Luca, and Naik}{Glaeser
  et~al.}{2018}]{glaeser2018big}
\textsc{Glaeser, E.~L., S.~D. Kominers, M.~Luca, and N.~Naik} (2018):
  \enquote{Big data and big cities: The promises and limitations of improved
  measures of urban life,} \emph{Economic Inquiry}, 56, 114--137.

\bibitem[\protect\citeauthoryear{Goodfellow, Bengio, and Courville}{Goodfellow
  et~al.}{2016}]{Goodfellow-et-al-2016}
\textsc{Goodfellow, I., Y.~Bengio, and A.~Courville} (2016): \emph{Deep
  Learning}, MIT Press.

\bibitem[\protect\citeauthoryear{Gross}{Gross}{2016}]{gross2016creativity}
\textsc{Gross, D.~P.} (2016): \enquote{Creativity under fire: The effects of
  competition on creative production,} \emph{Review of Economics and
  Statistics}, 1--17.

\bibitem[\protect\citeauthoryear{Hastings}{Hastings}{2004}]{hastings2004vertical}
\textsc{Hastings, J.~S.} (2004): \enquote{Vertical relationships and
  competition in retail gasoline markets: Empirical evidence from contract
  changes in Southern California,} \emph{American Economic Review}, 94,
  317--328.

\bibitem[\protect\citeauthoryear{He, Zhang, Ren, and Sun}{He
  et~al.}{2016}]{he2016deep}
\textsc{He, K., X.~Zhang, S.~Ren, and J.~Sun} (2016): \enquote{Deep residual
  learning for image recognition,} in \emph{Proceedings of the IEEE conference
  on computer vision and pattern recognition}, 770--778.

\bibitem[\protect\citeauthoryear{Hoberg and Phillips}{Hoberg and
  Phillips}{2016}]{hobergetal2016}
\textsc{Hoberg, G. and G.~Phillips} (2016): \enquote{Text Based Network
  Industries and Endogenous Product Differentiation,} \emph{Journal of
  Political Economy}, 124, 1423--1465.

\bibitem[\protect\citeauthoryear{Hotelling}{Hotelling}{1929}]{hotelling1929}
\textsc{Hotelling, H.} (1929): \enquote{Stability in Competition,}
  \emph{Economic Journal}, 39, 41--57.

\bibitem[\protect\citeauthoryear{Kovashka, Parikh, and Grauman}{Kovashka
  et~al.}{2012}]{kovashka2012whittlesearch}
\textsc{Kovashka, A., D.~Parikh, and K.~Grauman} (2012):
  \enquote{Whittlesearch: Image search with relative attribute feedback,} in
  \emph{2012 IEEE Conference on Computer Vision and Pattern Recognition}, IEEE,
  2973--2980.

\bibitem[\protect\citeauthoryear{Kozlowski, Taddy, and Evans}{Kozlowski
  et~al.}{2019}]{Kozlowski_2019}
\textsc{Kozlowski, A.~C., M.~Taddy, and J.~A. Evans} (2019): \enquote{The
  Geometry of Culture Analyzing the Meanings of Class through Word Embeddings,}
  \emph{American Sociological Review}, 84, 905--949.

\bibitem[\protect\citeauthoryear{Krizhevsky, Sutskever, and Hinton}{Krizhevsky
  et~al.}{2012}]{krizhevsky2012imagenet}
\textsc{Krizhevsky, A., I.~Sutskever, and G.~E. Hinton} (2012):
  \enquote{Imagenet classification with deep convolutional neural networks,} in
  \emph{Advances in neural information processing systems}, 1097--1105.

\bibitem[\protect\citeauthoryear{Lancaster}{Lancaster}{1971}]{lancaster1971consumer}
\textsc{Lancaster, K.} (1971): \emph{Consumer demand: A new approach}, Columbia
  University Press New York.

\bibitem[\protect\citeauthoryear{Lancaster}{Lancaster}{1966}]{lancaster1966new}
\textsc{Lancaster, K.~J.} (1966): \enquote{A new approach to consumer theory,}
  \emph{Journal of political economy}, 74, 132--157.

\bibitem[\protect\citeauthoryear{LeCun, Cortes, and Burges}{LeCun
  et~al.}{2010}]{lecun2010mnist}
\textsc{LeCun, Y., C.~Cortes, and C.~Burges} (2010): \enquote{MNIST handwritten
  digit database,} \emph{AT{\&}T Labs}.

\bibitem[\protect\citeauthoryear{Magnolfi, McClure, and Sorensen}{Magnolfi
  et~al.}{2022}]{magnolfi2022triplet}
\textsc{Magnolfi, L., J.~McClure, and A.~Sorensen} (2022): \enquote{Triplet
  Embeddings for Demand Estimation,} \emph{SSRN working paper}.

\bibitem[\protect\citeauthoryear{Mall, Matzen, Hariharan, Snavely, and
  Bala}{Mall et~al.}{2019}]{mall2019geostyle}
\textsc{Mall, U., K.~Matzen, B.~Hariharan, N.~Snavely, and K.~Bala} (2019):
  \enquote{Geostyle: Discovering fashion trends and events,} in
  \emph{Proceedings of the IEEE International Conference on Computer Vision},
  411--420.

\bibitem[\protect\citeauthoryear{Mankiw and Whinston}{Mankiw and
  Whinston}{1986}]{mankiw1986free}
\textsc{Mankiw, N.~G. and M.~D. Whinston} (1986): \enquote{Free entry and
  social inefficiency,} \emph{The RAND Journal of Economics}, 48--58.

\bibitem[\protect\citeauthoryear{Mazzeo, Seim, and Varela}{Mazzeo
  et~al.}{2018}]{mazzeo2018}
\textsc{Mazzeo, M.~J., K.~Seim, and M.~Varela} (2018): \enquote{The Welfare
  Consequences of Mergers with Endogenous Product Choice,} \emph{The Journal of
  Industrial Economics}, 66, 980--1016.

\bibitem[\protect\citeauthoryear{McFadden}{McFadden}{1973}]{mcfadden1973conditional}
\textsc{McFadden, D.} (1973): \emph{Conditional logit analysis of qualitative
  choice behavior}, Institute of Urban and Regional Development, University of
  California Berkeley.

\bibitem[\protect\citeauthoryear{Mikolov, Chen, Corrado, and Dean}{Mikolov
  et~al.}{2013}]{mikolov2013efficient}
\textsc{Mikolov, T., K.~Chen, G.~Corrado, and J.~Dean} (2013):
  \enquote{Efficient estimation of word representations in vector space,}
  \emph{arXiv preprint arXiv:1301.3781}.

\bibitem[\protect\citeauthoryear{Nevo}{Nevo}{2001}]{nevo2001measuring}
\textsc{Nevo, A.} (2001): \enquote{Measuring market power in the ready-to-eat
  cereal industry,} \emph{Econometrica}, 69, 307--342.

\bibitem[\protect\citeauthoryear{Nevo and Whinston}{Nevo and
  Whinston}{2010}]{nevo2010taking}
\textsc{Nevo, A. and M.~D. Whinston} (2010): \enquote{Taking the dogma out of
  econometrics: Structural modeling and credible inference,} \emph{Journal of
  Economic Perspectives}, 24, 69--82.

\bibitem[\protect\citeauthoryear{O'Donovan, L{\=\i}beks, Agarwala, and
  Hertzmann}{O'Donovan et~al.}{2014}]{o2014exploratory}
\textsc{O'Donovan, P., J.~L{\=\i}beks, A.~Agarwala, and A.~Hertzmann} (2014):
  \enquote{Exploratory font selection using crowdsourced attributes,} \emph{ACM
  Transactions on Graphics (TOG)}, 33, 1--9.

\bibitem[\protect\citeauthoryear{Parikh and Grauman}{Parikh and
  Grauman}{2011}]{parikh2011relative}
\textsc{Parikh, D. and K.~Grauman} (2011): \enquote{Relative attributes,} in
  \emph{2011 International Conference on Computer Vision}, IEEE, 503--510.

\bibitem[\protect\citeauthoryear{Rosen}{Rosen}{1974}]{rosen1974hedonic}
\textsc{Rosen, S.} (1974): \enquote{Hedonic prices and implicit markets:
  product differentiation in pure competition,} \emph{Journal of political
  economy}, 82, 34--55.

\bibitem[\protect\citeauthoryear{Schroff, Kalenichenko, and Philbin}{Schroff
  et~al.}{2015}]{schroff2015facenet}
\textsc{Schroff, F., D.~Kalenichenko, and J.~Philbin} (2015): \enquote{FaceNet:
  A unified embedding for face recognition and clustering,} in \emph{2015 IEEE
  Conference on Computer Vision and Pattern Recognition (CVPR)}, 815--823.

\bibitem[\protect\citeauthoryear{Seim}{Seim}{2006}]{siem2006spatial}
\textsc{Seim, K.} (2006): \enquote{An Empirical Model of Firm Entry with
  Endogenous Product-Type Choices,} \emph{The RAND Journal of Economics}, 37,
  619--640.

\bibitem[\protect\citeauthoryear{Simonyan and Zisserman}{Simonyan and
  Zisserman}{2014}]{simonyan2014very}
\textsc{Simonyan, K. and A.~Zisserman} (2014): \enquote{Very deep convolutional
  networks for large-scale image recognition,} \emph{arXiv preprint
  arXiv:1409.1556}.

\bibitem[\protect\citeauthoryear{Sun, Wang, and Tang}{Sun
  et~al.}{2015}]{sun2015deeply}
\textsc{Sun, Y., X.~Wang, and X.~Tang} (2015): \enquote{Deeply learned face
  representations are sparse, selective, and robust,} in \emph{Proceedings of
  the IEEE conference on computer vision and pattern recognition}, 2892--2900.

\bibitem[\protect\citeauthoryear{Sweeting}{Sweeting}{2010}]{sweeting2010effects}
\textsc{Sweeting, A.} (2010): \enquote{The effects of mergers on product
  positioning: evidence from the music radio industry,} \emph{The RAND Journal
  of Economics}, 41, 372--397.

\bibitem[\protect\citeauthoryear{Sweeting}{Sweeting}{2013}]{sweeting2013}
---\hspace{-.1pt}---\hspace{-.1pt}--- (2013): \enquote{Dynamic Product
  Positioning in Differentiated Product Markets: The Effect of Fees for Musical
  Performance Rights on the Commercial Radio Industry,} \emph{Econometrica},
  81, 1763--1803.

\bibitem[\protect\citeauthoryear{Taigman, Yang, Ranzato, and Wolf}{Taigman
  et~al.}{2014}]{taigman2014deepface}
\textsc{Taigman, Y., M.~Yang, M.~Ranzato, and L.~Wolf} (2014):
  \enquote{Deepface: Closing the gap to human-level performance in face
  verification,} in \emph{Proceedings of the IEEE conference on computer vision
  and pattern recognition}, 1701--1708.

\bibitem[\protect\citeauthoryear{Tenenbaum and Freeman}{Tenenbaum and
  Freeman}{2000}]{tenenbaum2000separating}
\textsc{Tenenbaum, J.~B. and W.~T. Freeman} (2000): \enquote{Separating style
  and content with bilinear models,} \emph{Neural computation}, 12, 1247--1283.

\bibitem[\protect\citeauthoryear{Wang, Song, Rosenberg, Wang, Chen, and
  Wu}{Wang et~al.}{2014}]{wang2014triplet}
\textsc{Wang, J., T.~Song, Yangand~Leung, C.~Rosenberg, J.~Wang,
  Jingbinand~Philbin, B.~Chen, and Y.~Wu} (2014): \enquote{Learning
  Fine-grained Image Similarity with Deep Ranking,} \emph{arXiv preprint
  arXiv:1404.4661}.

\bibitem[\protect\citeauthoryear{Wilson and Martinez}{Wilson and
  Martinez}{2003}]{wilson2003general}
\textsc{Wilson, D.~R. and T.~R. Martinez} (2003): \enquote{The general
  inefficiency of batch training for gradient descent learning,} \emph{Neural
  networks}, 16, 1429--1451.

\bibitem[\protect\citeauthoryear{Wollmann}{Wollmann}{2018}]{wollmann2018trucks}
\textsc{Wollmann, T.~G.} (2018): \enquote{Trucks without bailouts: Equilibrium
  product characteristics for commercial vehicles,} \emph{American Economic
  Review}, 108, 1364--1406.

\bibitem[\protect\citeauthoryear{Yu and Grauman}{Yu and
  Grauman}{2019}]{yu2019thinking}
\textsc{Yu, A. and K.~Grauman} (2019): \enquote{Thinking outside the pool:
  Active training image creation for relative attributes,} in \emph{Proceedings
  of the IEEE Conference on Computer Vision and Pattern Recognition}, 708--718.

\bibitem[\protect\citeauthoryear{Zhang, Lee, Singh, and Srinivasan}{Zhang
  et~al.}{2017}]{zhang2017much}
\textsc{Zhang, S., D.~D. Lee, P.~V. Singh, and K.~Srinivasan} (2017):
  \enquote{How much is an image worth? Airbnb property demand estimation
  leveraging large scale image analytics,} \emph{SSRN}.

\end{thebibliography}

\end{document}